# Collaborative Optimization Framework of Battery Charging / Swapping Stations for eVTOLs Based on Closed-Loop Supply Chain and Space-Time Network

Pengfeng Lin[1], Miao Zhu[1*], Jiahui Sun[2], Haoyang Cui[2], Chuanlin Zhang[2], Yunda Yan[3]

[1] School of Electrical Engineering, Shanghai Jiao Tong University, Shanghai 200240, China
[2] Faculty of Artificial Intelligence, Shanghai University of Electric Power, Shanghai 201306, China
[3] Department of Computer Science, University College London, London WC1E 6BT, UK
[*]miaozhu@sjtu.edu.cn

**Abstract:** Electric vertical take-off and landing (eVTOL) aircraft, by virtue of their advantages such as high efficiency, environmental friendliness, and low noise, are regarded as an effective means of alleviating urban traffic congestion. Although eVTOLs have policy support, industrial foundations, and market prospects globally, their commercialization process still faces obstacles. Purely electric eVTOLs confront constraints including limited battery energy density, high operational power requirements, and challenges associated with rapid energy replenishment, which collectively restrict their flight endurance and application scenarios. To tackle these issues, following an investigation into battery energy replenishment strategies, a closed-loop supply chain-based optimization framework for eVTOL battery charging and swapping is proposed. First, a time-space network technique is employed to solve the modeling challenge of an efficient and flexible battery transportation system between charging stations and swapping stations. Subsequently, aiming to maximize the operational revenue of the model, optimized management of battery swapping, transportation, and charging processes is implemented, facilitating coordinated operation among eVTOLs, swapping stations, and charging stations. On this basis, the model is solved by Gurobi, verifying its feasibility. Finally, the results of sensitivity analysis show that the model performs excellently in several aspects, including battery swap service quality, system sustainability, and operational resilience.

## I. INTRODUCTION

With the continuous advancement of urbanization, the global urban population continues to grow and is projected to account for 68% of the global total population by 2050 [1]. Although urbanization brings many benefits, it also leads to traffic congestion, which in turn causes economic losses. According to the INRIX congestion report [2], the national congestion loss in The United States reached USD 81 billion in 2022, representing a 53% increase from the previous year, with major cities such as New York being the most severely affected. The underlying cause of this situation is the contradiction between continuously growing travel demand and limited urban road capacity. Therefore, if urban transportation could be extended into three-dimensional space, it is expected to alleviate traffic congestion.

Against this background, urban air mobility (UAM), which provides transportation services over urban areas, has emerged as an effective means of alleviating congestion [3]. Among various aircraft, electric vertical take-off and landing (eVTOL) has become the core vehicle for UAM mainly due to the following five advantages. First, eVTOL can fly flexibly over urban areas, not only avoiding traffic congestion but also enabling faster point-to-point direct travel [4], while simultaneously relieving some of the pressure on ground transportation and improving the efficiency of the overall transportation system [5]. Second, it is powered by batteries and driven by electric motors, producing no air pollution during flight [4]. Third, compared to conventional helicopters, eVTOL has extremely low noise levels and does not cause disturbance to residents. Fourth, even when some of its rotors fail, it can still maintain stable flight, offering a high level of safety [1]. Fifth, it has strong spatial adaptability, requiring only small vertiports or urban rooftop platforms for takeoff and landing [6], thus imposing low demands on ground infrastructure. Owing to the above advantages, eVTOL has a wide range of application scenarios within cities, covering areas such as intra-city and inter-city transportation [7], medical rescue [8-9], and tourism [10].

To promote the application and commercialization of eVTOL, various governments have successively adopted supportive measures. The European Union released "NPA 2022-06" in 2022, which contains regulations related to UAM certification and operations [11], playing a leading role in the global legal domain. In 2023, the Chinese government designated the low-altitude economy as one of the key strategic emerging industries [12], not only granting the low-altitude economy a relatively high strategic status but also effectively promoting the development of the eVTOL industry. In 2024, the UK Department for Transport published the "Future of Flight Action Plan," clarifying its future development roadmap for eVTOL [13].

In addition to government efforts, various industries have also entered the eVTOL market, including but not limited to emerging eVTOL companies, traditional aviation giants, and automotive enterprises. For example, EHang's EH216-S obtained the world's first type certificate for eVTOL [8] and has since received orders from multiple countries including Japan, Singapore, and Indonesia [14]. According to forecasts by institutions such as MarketsandMarkets and Morgan Stanley [15-16], the market value of eVTOL will exceed USD 23.4 billion by 2030, the total output value of UAM could reach USD 1 trillion by 2040, and China is expected to become the world's largest eVTOL market. However, despite strong policy support, a solid industrial foundation, and a broad blue ocean market, the commercialization path of eVTOL still faces challenges.

Insufficient battery endurance is a major challenge faced by eVTOL, severely restricting its range of applications. Specifically, battery-powered eVTOL is primarily plagued by issues such as low battery energy density [3,8], high operational power requirements [17-18], and difficulties in fast charging [5,19]. Currently, the energy density of the most advanced single battery cell is approximately 300 Wh/kg, which represents a substantial gap compared to the 10,000–12,000 Wh/kg of conventional aviation fuel, limiting eVTOL to short-distance flights. Meanwhile, compared to new energy vehicles(EV), eVTOL has an extremely high power demand during the vertical take-off and landing phase, requiring the battery discharge rate to reach at least 4–5C, and its peak power duration is longer, further restricting endurance performance. Furthermore, to achieve profitability, operators typically schedule eVTOL for high-frequency takeoffs and landings, making fast charging during brief stopovers critically important. However, current fast-charging technology for eVTOL remains immature and cannot achieve deep battery recharging in a short time, directly reducing operational efficiency. Therefore, there is an urgent need to investigate fast battery energy replenishment solutions for eVTOL.

eVTOL has two energy replenishment strategies: plug-in charging and battery swapping [20]. For plug-in charging, due to the immaturity of fast-charging technology, even if operators allow eVTOL to charge for extended periods at vertiports, it may still lead to congestion at charging piles and parking stands, hindering the landing and energy replenishment of subsequent eVTOL [21]. Furthermore, batteries subjected to long-term fast charging have a lower thermal runaway triggering temperature, and the continuous temperature rise during fast charging can easily induce safety issues such as thermal runaway [22-23]. Reference [17] further indicates that increasing the charging rate accelerates battery aging and shortens its cycle life. Therefore, under current practical constraints, the battery swapping strategy becomes a more favorable option. First, the swapping process takes very little time; eVTOL can depart immediately after battery swapping and passenger transfer, thereby reducing vertiport occupancy while maintaining high operational efficiency. Second, batteries that have just completed a flight mission are at elevated temperatures; removing them allows sufficient time for cooling before being placed into chargers for slow charging. This approach not only avoids safety issues but also helps slow down the degradation of battery capacity and lifespan [19]. Finally, the study in [20] confirms that the swapping strategy is more conducive to reducing operational costs.

There is an operational model for battery swapping [24], in which a battery swapping station (BSS) undertakes all operations including charging, swapping, and ancillary services. However, this model also has certain drawbacks that cannot be ignored. First, to facilitate convenient access to swapping services for users, BSSs are typically located in densely populated areas; however, storing and charging a large number of batteries in such areas poses serious safety risks to surrounding personnel [25]. Second, the BSS undertakes all operations, which imposes substantial land area requirements and further complicates its internal structure [26]. Third, the BSS must bear various costs such as battery procurement and land acquisition [27], leading to enormous upfront construction costs. Finally, the transmission capacity of grid transformers in many old urban areas is limited, making it difficult to withstand the load impact of large-scale centralized charging. Therefore, some studies have proposed an improved battery charging and swapping system model (BCSS) [25], which effectively overcomes the limitations of the BSS by separating charging and swapping services. Under this model, the battery charging station (BCS) is responsible for centralized battery charging, while the BSS only needs to focus on swapping services. In this case, existing infrastructure such as parking structures and building rooftops can serve as BSSs, allowing them to integrate into the urban transportation system without increasing pressure on land resources, thereby reducing upfront construction costs.

Based on the above, this paper introduces the BCSS into the study of fast energy replenishment strategies for eVTOL and draws two insights from relevant research in the EV domain. First, the battery logistics system between the BCS and BSS is critical to ensuring the smooth operation of the energy replenishment system and therefore requires focused modeling. Second, the BCSS involves multiple entities, including the BCS, BSS, and transportation fleet; optimization of the system should balance the interests of all parties and strengthen coordination among them. The closed-loop supply chain (CLSC) is a management model that achieves resource recycling [28] and is both generalizable and capable of generating multiple benefits. Accordingly, this paper constructs a BCSS for eVTOL based on its theoretical framework. However, this paper must also establish an efficiently integrated logistics transportation system and ensure that the system maintains high flexibility over the long term; otherwise, the BSS may have no batteries available for swapping and the BCS may have no batteries to charge, potentially leading to operational disruption. Given that time-space network (TSN) techniques facilitate modeling and solution when dealing with timing-dependent problems [29], this paper adopts TSN to address the challenge of constructing the logistics transportation system.

In summary, this paper proposes an eVTOL battery charging and swapping system model based on CLSC and TSN. On this basis, subject to satisfying the battery swapping demand of eVTOLs, the battery swapping, transportation, and charging processes are jointly optimized, thereby achieving coordinated operation among eVTOLs, BSSs, and BCSs while maximizing the overall operational revenue of the system. Finally, the feasibility and superiority of the proposed model are verified through simulation.

## II. Technical review of Related Studies

Before conducting research on battery charging and swapping strategies for eVTOL, inspiration can be drawn from relevant studies on EVs. Although the two differ in terms of operating environments, they share many common research objectives in the area of battery energy replenishment, such as optimizing charging scheduling, improving operational revenue, and interacting with the power grid. Furthermore, since CLSC and TSN serve as the core foundation and key technologies of this paper, it is necessary to review the relevant research on both. To this end, this paper conducts a literature survey focusing on the following six areas: (A) research on battery swapping strategies for EVs, (B) research on battery charging and swapping strategies for EVs, (C) research on plug-in charging strategies for eVTOLs, (D) research on charging and swapping strategies for eVTOLs, (E) research on CLSC, and (F) research on TSN.

### A. *Research on battery swapping strategies for EVs*

In terms of operational optimization, reference [30] proposes a dynamic scheduling method for BSS based on rolling horizon optimization, which utilizes heterogeneous battery inventory and continuously controllable chargers to provide optimal battery swapping services and charging/discharging schedules for different types of EVs and batteries, thereby increasing the number of EVs served and daily profit. In terms of resource allocation, reference [31] establishes an optimal configuration model for BSS with the objective of minimizing investment, maintenance, and electricity costs, obtaining quantitative relationships among EVs, batteries, and chargers under different waiting times, which provides a reference for core asset allocation in BSS. In terms of macro-level planning, reference [32] employs the proposed algorithm to solve the problem of simultaneously determining the optimal siting of BSS and the optimal routing of EV fleets. Regarding interaction with the power grid, given the randomness of battery charging load demand in BSS, reference [33] conducts research on this issue and proposes charging guidance, effectively mitigating the impact of disorderly charging behavior on the distribution network.

### B. *Research on battery charging and swapping strategies for EVs*

Reference [34] systematically proposed the concept of BCSS at an early stage and studied the operational mechanism of EV battery charging and swapping service networks, although it did not take the transportation network into consideration. Reference [35], on the other hand, focused on the battery scheduling optimization problem between BCS and BSS, constructing a scheduling model aimed at minimizing transportation costs, but it did not fully address the operational management issues of the two entities. Over time, research on BCSS has been gradually refined and expanded. Reference [36] explored the real-time optimal scheduling problem of BCSS, incorporating both battery charging and swapping and truck path planning into the decision-making framework, and proposed an algorithm combining deep reinforcement learning with binary integer programming to address dynamic hard constraints in the scheduling process. Reference [37] established a planning framework for BCSS aimed at minimizing investment and operational costs. On one hand, the framework determines the optimal siting and sizing of BSSs with the objective of reducing BSS investment costs and battery transportation costs. On the other hand, it provides branch reinforcement schemes for the distribution network to minimize network reinforcement and energy loss costs.

### C. *Research on plug-in charging strategies for eVTOLs*

Several scholars have conducted multiple studies on scheduling optimization for eVTOL plug-in charging based on different optimization objectives. These works have not only focused on the charging behavior itself but have also integrated it with fleet scheduling, service quality, and grid impact. The scheduling method in reference [38] comprehensively considers battery state of charge, charging station capacity, and logistics demand, optimizing aircraft routes and charging schedules in real time, which not only reduces the total fleet travel time but also significantly improves charging facility utilization. The model proposed in reference [39] can solve the joint optimization problem of charging scheduling and passenger service within a short computation time, satisfying eVTOL charging demand while minimizing total passenger waiting time. However, one prerequisite of this approach is that the vertiport must have sufficient space for eVTOL to wait for charging. As the scale of eVTOL applications continues to expand, their charging demand will have a significant impact on the power grid, and some studies have incorporated the grid into the scheduling decision framework. Reference [40] uses distribution network power flow simulation to find that large-scale eVTOL charging can cause power quality issues, necessitating upgrades to existing grid infrastructure and the deployment of voltage management strategies to meet daily operational demands. Reference [41] first generates a feasible charging plan based on estimated passenger travel demand, then, without affecting eVTOL scheduling, intelligently optimizes charging timing and rates in conjunction with the dynamic characteristics of the grid to maximize power system benefits. Therefore, in addition to the drawbacks of occupying vertiport land, reducing battery capacity and lifespan, and posing safety risks, establishing a BCS capable of supporting simultaneous fast charging for multiple eVTOL entails extremely high construction and grid upgrade costs, while the potential adverse impacts on the grid cannot be ignored.

### D. *Research on charging and swapping strategies for eVTOLs*

Given the current limitations of plug-in charging, some studies have begun to explore charging and swapping strategies for eVTOL and have confirmed the feasibility and superiority of such hybrid modes. Reference [20] designed two energy replenishment strategies, plug-in charging and battery swapping, for electric aircraft within an airport microgrid, and conducted a comparative analysis from both economic and technical dimensions. The results indicate that the swapping strategy is superior in reducing operational costs. Reference [42] analyzed the operational results of the proposed model and found that combining charging and swapping strategies can triple the workload executable by the fleet. Reference [19] improved upon existing scheduling algorithms and proposed power optimization strategies and power investment optimization strategies for battery swapping and charging. The former aims to minimize peak power drawn from the grid and electricity costs, while the latter seeks to minimize electricity bills and capital expenditures related to battery and charger procurement. However, these studies share a common characteristic: they rely on large airports capable of accommodating both BCS and BSS as takeoff and landing sites for eVTOL. Such a setup is not problematic in itself, but there remains room for improvement. The reason is that if the goal is to alleviate traffic congestion in large cities through eVTOL applications, the improvement achievable by dispatching eVTOL only between airports is very limited. Existing urban infrastructure such as parking structures or building rooftops could also be fully utilized, and new vertiports could even be constructed. However, integrating both battery swapping and centralized charging at these locations clearly faces significant difficulties, implying not only high upfront construction investments but also exacerbating the scarcity of urban land resources, along with other constraints. Therefore, if the functionality of takeoff and landing sites is simplified to provide only battery swapping and passenger transfer services, more small vertiports can be established with minimal resources, thereby making it easier to integrate eVTOL into existing urban transportation systems.

### E. *Research on CLSC*

CLSC is an advanced management model that can unlock the value potential of products throughout their life cycle while minimizing energy consumption and environmental pollution [43], thereby promoting sustainable economic development. In the field of retired EV battery recycling, reference [44] constructs four hybrid channel recycling models based on CLSC and derives the optimal pricing strategy and profit maximization scheme for each CLSC member, ultimately identifying the best recycling model for each specific scenario. In the field of logistics transportation, when manufacturers face challenges arising from uncertain demand from retailers and random requests for picking up used products, reference [45] investigates the fleet sizing and vehicle routing problem in a CLSC considering delivery and pickup, and constructs a model that minimizes vehicle procurement and transportation costs. In the eVTOL battery energy replenishment system, when depleted batteries (DBs) are removed at a BSS, transported to a BCS for charging, and then returned as well-charged batteries (WBs) to the BSS, this closed-loop cycle perfectly aligns with the operational mechanism of a CLSC, not only improving the utilization rate of existing batteries but also reducing the expenditure on purchasing new batteries. Therefore, based on the existing theoretical framework of CLSC, this paper constructs a BCSS model for eVTOLs.

### F. Research on TSN

TSN, which offers advantages such as ease of modeling and solution when dealing with timing-dependent problems, has been applied in operational optimization studies in fields such as power systems. For example, reference [46] constructs a TSN model related to the transportation network and distribution system to address the joint post-disaster recovery and network reconfiguration problem of mobile energy storage systems and generation dispatch in microgrids. On this basis, by reducing customer outage costs as well as generation and energy storage costs, the post-disaster recovery capability and resilience of the distribution system are enhanced. In the context of congestion management in coupled transportation and power systems, reference [47] proposes a TSN model with queuing arcs to describe the spatiotemporal distribution of EVs in road networks and BCSs, clearly presenting capacity and storage constraints for roads and charging stations, as well as continuous charging and discharging strategies. In the eVTOL BCSS model proposed in this paper, BSSs and BCSs are independent and geographically non-adjacent, necessitating the establishment of a logistics transportation system between them, and this system must possess high flexibility. To address the challenge of constructing the logistics transportation system and ensure operational continuity of the entire BCSS model, this paper employs TSN to model it.

## III. Modeling and Operational Principles of the eVTOL Battery Charging and Swapping System

This section selects a suitable CLSC model by analyzing the characteristics of each member in the BCSS. Subsequently, a model of the eVTOL battery charging and swapping system based on the selected CLSC model is constructed, and its working principle is elaborated.

### A. Closed-Loop Supply Chain Model Selection

In the research scenario of this paper, the energy replenishment scheme for eVTOL batteries is characterized by swapping first and charging later, with swapping and charging operations decoupled. Specifically, eVTOL replaces DBs with WBs at a BSS and pays a service fee after the swap. The BCS receives DBs from BSSs in batches for centralized charging while supplying an adequate number of WBs to the BSSs. Since the BCS and BSSs are geographically non-adjacent, the transfer of DBs and WBs between them relies on a dedicated transportation fleet. Thus, the BCSS constructed in this paper encompasses four major members: eVTOL, BSS, BCS, and the transportation fleet, where eVTOL, WB, DB, BSS, and BCS can be regarded as consumers, new products, used products, retailers, and manufacturers, respectively.

After analyzing the functions of each member in the BCSS, three trade-in CLSC models[28] are available for selection. In Model R, retailers wholesale new products from manufacturers and sell them to consumers, while used products are returned by consumers to retailers and then sent back to the manufacturers. In Model C, manufacturers and retailers are integrated into a single enterprise, selling products directly to consumers while simultaneously recovering used products. Model M differs from Model R in that used products are recovered directly by manufacturers from consumers. In the BCSS, Model C is suitable for scenarios where the BSS can handle all operations, while Model M requires that eVTOL first swap batteries at a BSS and then send the DBs to the BCS for charging, an arrangement that is clearly unreasonable in practice. Since the scenario described in this paper does not align with the conditions of the two aforementioned models, Model R is adopted as the CLSC model.

*B. Working Principle of the eVTOL Battery Charging and Swapping System Model Based on CLSC*

The overall structure of the BCSS based on Model R proposed in this paper is shown in Fig. 1. For convenience of discussion, this paper establishes eight assumptions regarding the proposed BCSS, described in detail as follows:

(1) In actual operations, BCSS needs to serve multiple models of eVTOLs; however, their batteries exhibit significant differences in parameters such as volume, energy density, and maximum acceptable charging power, posing enormous challenges to the centralized charging and management of batteries. To reduce the technical complexity introduced by this heterogeneity, this paper stipulates that BCSS serves only Airbus's CityAirbus, uniformly adopting dedicated batteries matched to its operational requirements.

(2) The energy consumption of eVTOL batteries is influenced by the coupling effects of multiple factors such as flight altitude, passenger load, and meteorological conditions, making it difficult to ensure that all batteries are exactly depleted upon arrival at the BSS. To simplify model complexity, this paper stipulates that all DBs with a state of charge not exceeding the minimum state of charge $SOC_{\min}$ must be replaced at the BSS; meanwhile, all WBs provided by the BSS to eVTOLs are at the maximum state of charge $SOC_{\max}$.

(3) The BCS in the system purchases electricity from the power grid at prices set by the grid operator.

(4) The BCS is equipped with a certain number of charging piles, but this number is less than the total quantity of batteries in the BCSS. Once fully charged, batteries must be removed from the charging piles and stored in the BCS warehouse. Additionally, the BCS reserves space for storing DBs.

(5) The BCS can lease backup DBs from a third-party company at a cost if its battery supply is insufficient.

(6) When the inventory of DBs at the BCS exceeds the number of available charging piles, the BCS may directly select DBs from the inventory for charging in subsequent time steps, without the need to lease spare DBs. Furthermore, this paper focuses solely on the macroscopic impact of the DB inventory quantity, the number of DBs retrieved, and the remaining quantity on the subsequent operation of the BCSS; therefore, no rule has been established for "precisely which DB to select from the inventory for charging," nor has any constraint been formulated that "DBs returned to inventory earlier must be charged with priority."

(7) The total supply of WBs from the BCS can meet the battery swapping service demand of all BSSs.

(8) Storing batteries on a large scale at a BSS significantly increases construction costs and introduces potential safety hazards; however, fully implementing a "zero-inventory" strategy makes it difficult to meet the battery swapping demand during the initial operational phase as well as sudden peak swapping demands. Therefore, to ensure the battery swapping service quality of the BSS, this paper stipulates that it must reserve an appropriate quantity of initial WB and DB inventory.

Taking the actual geographical conditions of Shanghai as an example, the working principle of the BCSS is introduced in detail in the form of daily operations. This paper assumes that three BSSs, one BCS, and a transportation fleet consisting of four trucks are deployed within the city. Thanks to the initial WB inventory, the BSS on Chongming Island can provide immediate battery swapping services to arriving eVTOLs without waiting for incoming trucks. At the BSS in downtown Shanghai, a truck is unloading WBs and loading DBs while the station simultaneously performs battery swaps for an eVTOL. At the BSS located at the border of Qingpu District and Songjiang District, following replenishment and clearance by trucks, the WB inventory is sufficient, and only DBs generated from the current swap operation remain on-site. Meanwhile, the BCS in Fengxian District, serving as the system's charging hub, is simultaneously handling battery charging, truck loading, and unloading tasks to ensure the stable operation of the battery logistics system.

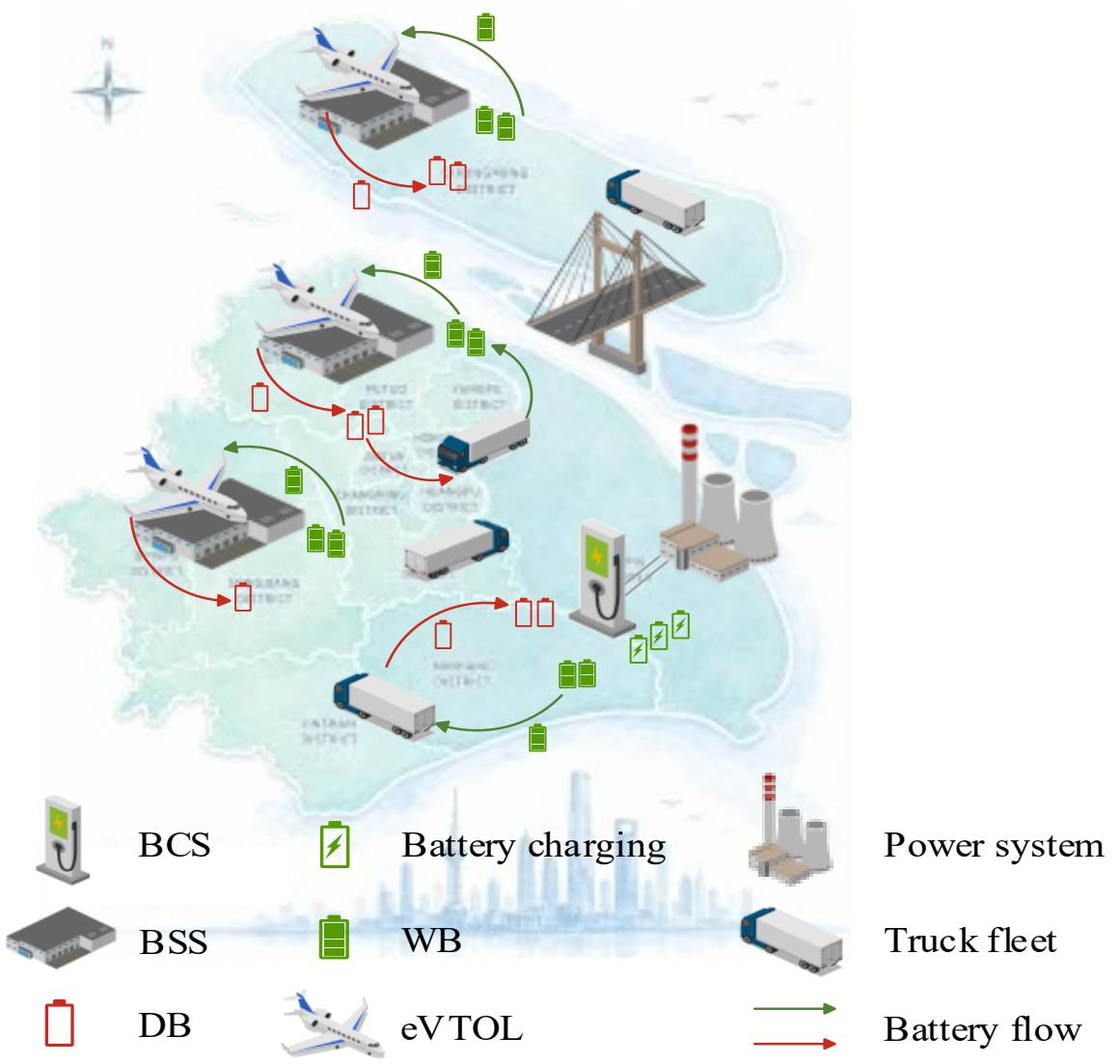

Fig. 1. A CLSC-based BCSS for eVTOL(taking Shanghai city as an example).

## IV. Optimal Operation Management for the CLSC-based eVTOL Battery Charging and Swapping System

Based on the BCSS described in the previous section and the scenario illustrated in Fig. 1, this section establishes operational models for the logistics system, BSSs, and the BCS, respectively, and presents their corresponding objective functions. Finally, these are integrated into the overall objective function of the BCSS proposed in this paper.

### *A. Logistics System Operation Model Based on TSN*

This paper constructs a BCSS model that integrates eVTOL battery operations, urban traffic dispatch, and power system regulation. To address the scheduling optimization problem for trucks and eVTOL batteries within this system, this section proposes a logistics system operational model based on TSN.

#### *1) Vehicle Scheduling and Battery Scheduling Models*

As shown in Fig. 2(a), the BCSS comprises one BCS and three BSSs, corresponding to nodes 1-4 respectively, and is interconnected via a transportation network. Each edge in the network is labeled with the number of time steps required for a truck to travel from one node to another. For instance, traveling from the BCS to BSS1 requires one time step. For nodes in the network that are connected but have a travel time greater than one time step, introducing virtual nodes between them ensures that the distance between all nodes in the network is one time step, thereby guaranteeing a defined state for each edge at every time step. As illustrated in Fig. 2(b), this paper introduces a virtual node5 between the BCS and BSS2.

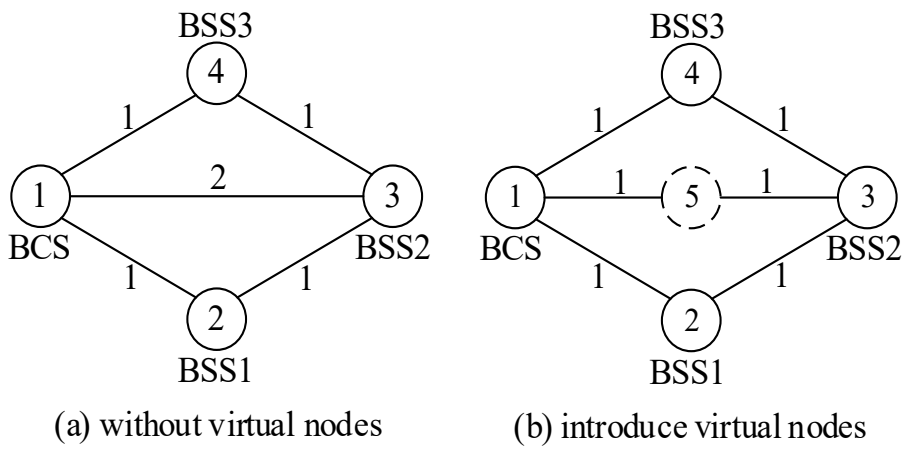

Fig. 2. Transportation network of BCSS.

In Fig. 3, the transportation network is modeled as a TSN, consisting of time steps, nodes, and arcs representing travel paths. Two types of arcs are used to represent truck parking and travel behaviors, respectively. For instance, an arc connecting the same node across consecutive time steps is a parking arc, indicating the truck is parked at that node. An arc connecting different nodes across consecutive time steps is a travel arc, indicating the truck is traveling on that route. Furthermore, there are no parking arcs at the virtual node. Using the TSN, the BCSS can track the state, location, and task of any truck at each time step, thereby enabling optimal scheduling of the logistics system based on the battery swapping demand of eVTOLs.

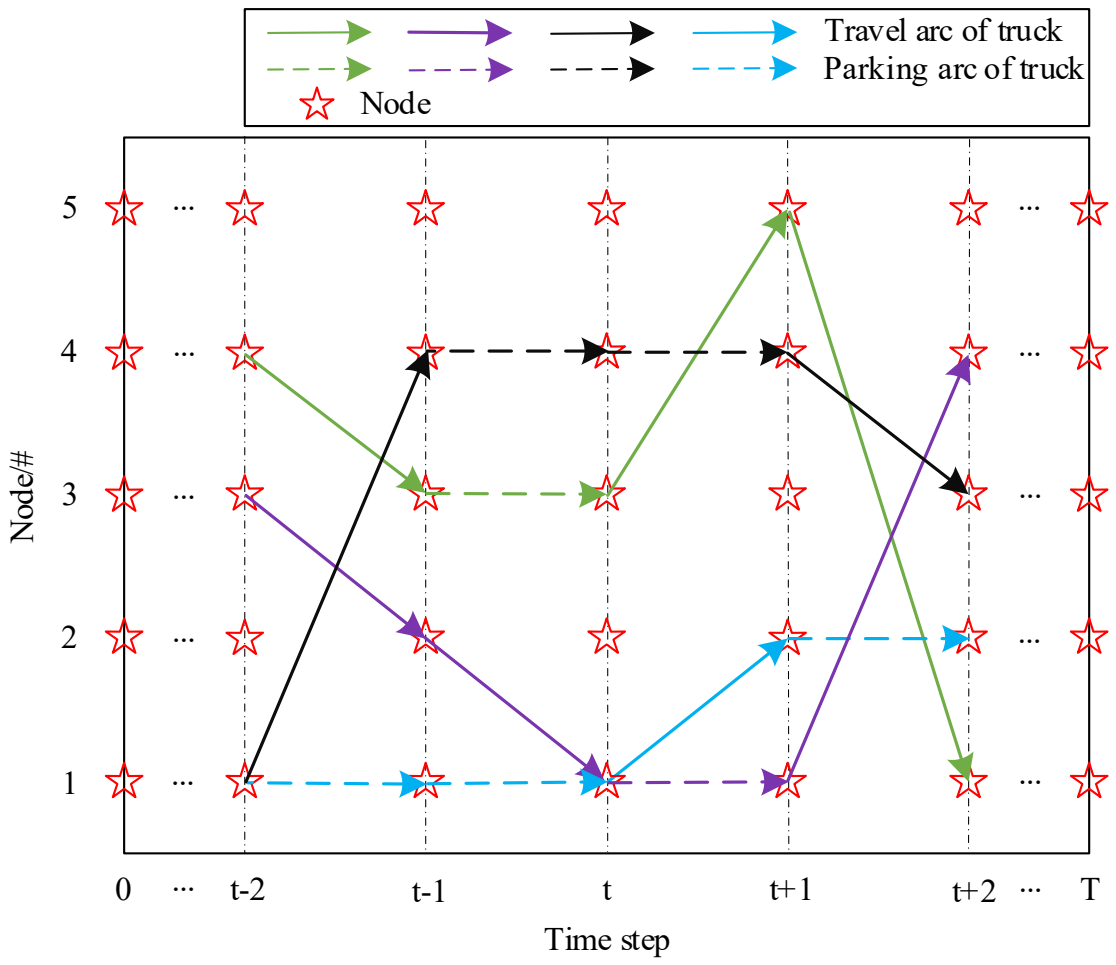

Fig. 3. Logistics system operation model based on TSN.

The operational model for the logistics system proposed in this paper comprises a vehicle scheduling model and a battery scheduling model for each truck. The vehicle scheduling model is presented in Equations (1)-(4). The real-time status of each truck serves as a critical basis for the BCSS to assign tasks for the next time step. Therefore, it must be ensured that a truck can only be in either a "traveling" or "parked" state at any time step:

$$\sum_{(i,j)\in Z} \zeta_{\mathrm{arc},(i,j)}^{\omega,t} = 1, \forall \omega \in H, t \in \mathcal{T} \tag{1}$$

where $\zeta_{\mathrm{arc},(i,j)}^{\omega,t}$ is a binary variable, if truck $\omega$ is on the arc starting from node $i$ and heading to node $j$ at time step $t$, the value of this variable is 1, $Z$ is the set of all arcs in the TSN, $(i,j)$ represents the arc with $i$ as the start node and $j$ as the end node, and is an element of $Z$, $H$ is the set of all trucks, $\omega$ represents the identification number of each truck, and is an element of $H$, $\mathcal{T}$ is the set of all time steps, $t$ denotes each time step, and is an element of $\mathcal{T}$.

This paper sets that truck $\omega$ is located at node $i$ at the initial moment and can only depart from this node at time step1. Therefore, we have:

$$\sum_{(i,j)\in Z_i^+} \zeta_{\mathrm{arc},(i,j)}^{\omega,1} = \zeta_{\mathrm{arc},i}^{\omega,\mathrm{ini}}, \forall \omega \in H, i,j \in G_{\mathrm{R}} \cup G_{\mathrm{V}} \tag{2}$$

where $Z_i^+$ is the set of arcs starting from node $i$, $\zeta_{\mathrm{arc},i}^{\omega,\mathrm{ini}}$ is a binary variable indicating that truck $\omega$ is located at node $i$ at the initial time step, "ini" is an abbreviation for "initial", indicating the initial time step, $G_{\mathrm{R}}$ is the set of actual nodes, $G_{\mathrm{V}}$ is the set of virtual nodes.

To prevent trucks from disappearing or appearing arbitrarily, continuity must be ensured in their scheduling between nodes. If the destination node of a truck at time step $t$ is $j$, then its departure node at time step $t+1$ must also be $j$:

$$\begin{gathered}\sum_{(i,j)\in Z_j^-} \zeta_{\mathrm{arc},(i,j)}^{\omega,t} = \sum_{(j,k)\in Z_j^+} \zeta_{\mathrm{arc},(j,k)}^{\omega,t+1}, \forall \omega \in H, \\ i,j,k \in G_{\mathrm{R}} \cup G_{\mathrm{V}}, t \in \mathcal{T} \setminus \{T\}\end{gathered} \tag{3}$$

where $Z_j^-$ is the set of arcs ending at node $j$, $T$ represents the last time step.

However, to accurately calculate the total scheduling cost of the logistics system and further optimize the scheduling strategy, the BCSS must also track the specific state of each truck. Therefore, the truck transportation state variable is defined:

$$\beta_{\mathrm{tru_on}}^{\omega,t} = \sum_{(i,j)\in Z, i\neq j} \zeta_{\mathrm{arc},(i,j)}^{\omega,t}, \forall \omega \in H, i,j \in G_{\mathrm{R}} \cup G_{\mathrm{V}}, t \in \mathcal{T} \tag{4}$$

where $\beta_{\mathrm{tru_on}}^{\omega,t}$ is a binary variable, if the departure node and the destination node of truck $\omega$ at time step $t$ are different, this variable equals 1, indicating the truck is in a traveling state, otherwise, it equals 0.

Building upon the vehicle scheduling model, the battery scheduling model for each truck is jointly defined by Equations (5)-(15). Specifically, the expression for the quantity of DBs carried by each truck at each time step is:

$$S_{\mathrm{tru,DB,stor}}^{\omega,t} = \begin{cases} \sum_{n\in G_{\mathrm{BSS}}} N_{\mathrm{tru,BSS,DB}}^{\omega,n,t} - \sum_{m\in G_{\mathrm{BCS}}} N_{\mathrm{tru,BCS,DB}}^{\omega,m,t}, \forall \omega \in H, t=1 \\ S_{\mathrm{tru,DB,stor}}^{\omega,t-1} + \sum_{n\in G_{\mathrm{BSS}}} N_{\mathrm{tru,BSS,DB}}^{\omega,n,t} - \sum_{m\in G_{\mathrm{BCS}}} N_{\mathrm{tru,BCS,DB}}^{\omega,m,t}, \\ \forall \omega \in H, t \in \mathcal{T} \setminus \{1\} \end{cases} \tag{5}$$

where $S_{\mathrm{tru,DB,stor}}^{\omega,t}$ is the number of DBs carried by the truck $\omega$ at time step $t$, $N_{\mathrm{tru,BSS,DB}}^{\omega,n,t}$ is the number of DBs loaded at BSS $n$, $N_{\mathrm{tru,BCS,DB}}^{\omega,m,t}$ is the number of DBs unloaded at BCS $m$, $G_{\mathrm{BSS}}$ is the set of BSSs, $n$ is the index of its elements, $G_{\mathrm{BCS}}$ is the set of BCSs, $m$ is the index of its elements.

Similarly, the expression for the number of WBs, denoted as $S_{\mathrm{tru,WB,stor}}^{\omega,t}$, carried by the truck $\omega$ at each time step is:

$$S_{\mathrm{tru,WB,stor}}^{\omega,t} = \begin{cases} \sum_{m\in G_{\mathrm{BCS}}} N_{\mathrm{tru,BCS,WB}}^{\omega,m,t} - \sum_{n\in G_{\mathrm{BSS}}} N_{\mathrm{tru,BSS,WB}}^{\omega,n,t}, \forall \omega \in H, t=1 \\ S_{\mathrm{tru,WB,stor}}^{\omega,t-1} + \sum_{m\in G_{\mathrm{BCS}}} N_{\mathrm{tru,BCS,WB}}^{\omega,m,t} - \sum_{n\in G_{\mathrm{BSS}}} N_{\mathrm{tru,BSS,WB}}^{\omega,n,t}, \\ \forall \omega \in H, t \in \mathcal{T} \setminus \{1\} \end{cases} \tag{6}$$

where $N_{\text{tru,BCS,WB}}^{\omega,m,t}$ is the number of WBs loaded by the truck $\omega$ at BCS $m$ at time step $t$, $N_{\text{tru,BSS,WB}}^{\omega,n,t}$ is the number of WBs unloaded by the truck $\omega$ at BSS $n$ at time step $t$.

For the decision variables defined in Equations (5) and (6), their value ranges must be constrained to ensure that they conform to physical meaning and actual operating conditions. For example, the sum of the numbers of DBs and WBs carried by a truck at any time step shall not exceed its maximum transport capacity:

$$S_{\text{tru,DB,stor}}^{\omega,t} + S_{\text{tru,WB,stor}}^{\omega,t} \le C_{\text{tru,cap}}, \forall \omega \in H, t \in \mathcal{T} \tag{7}$$

where $C_{\text{tru,cap}}$ is the maximum number of batteries that can be carried by a truck.

In addition to the total capacity constraint shown in Equation (7), the numbers of DBs and WBs carried by a truck must also be constrained individually. The first is a basic constraint, i.e., both quantities must be non-negative integers and always remain within a reasonable range. Furthermore, this paper further stipulates that after time step $T$, the remaining numbers of WBs and DBs on each truck shall not exceed one-third of its maximum capacity. The above two constraints are expressed by Equations (8) and (9), respectively:

$$\left.\begin{array}{l} 0 \le S_{\text{tru,DB,stor}}^{\omega,t} \le C_{\text{tru,cap}}, \forall \omega \in H, t \in \mathcal{T} \\ 0 \le S_{\text{tru,WB,stor}}^{\omega,t} \le C_{\text{tru,cap}}, \forall \omega \in H, t \in \mathcal{T} \end{array}\right\} \tag{8}$$

$$\left.\begin{array}{l} S_{\text{tru,DB,stor}}^{\omega,T} \le C_{\text{tru,cap}} / 3, \forall \omega \in H \\ S_{\text{tru,WB,stor}}^{\omega,T} \le C_{\text{tru,cap}} / 3, \forall \omega \in H \end{array}\right\} \tag{9}$$

The prerequisite for a truck to load DBs at a BSS is that it must be docked at that BSS, and the loading quantity cannot exceed its transport capacity:

$$0 \le N_{\text{tru,BSS,DB}}^{\omega,n,t} \le C_{\text{tru,cap}} \zeta_{\text{arc},(n,n)}^{\omega,t}, \forall \omega \in H, \forall n \in G_{\text{BSS}}, t \in \mathcal{T} \tag{10}$$

where $\zeta_{\text{arc},(n,n)}^{\omega,t}$ is a binary decision variable indicating whether truck $\omega$ is docked at BSS $n$ at time step $t$, if the truck is docked at BSS $n$ at that time step, the variable takes the value of 1, and the number of DBs that can be loaded there can seek an optimal value between 0-$C_{\text{tru,cap}}$, if not docked, the variable takes the value of 0, meaning the truck cannot load DBs there.

When a truck unloads WBs at a BSS, it must adhere to the following constraints. Equation (11) requires that unloading can only occur when the truck is parked. Equation (12) stipulates that no WBs can be unloaded at time step 1, and at other time steps, the number unloaded cannot exceed the quantity of WBs carried by the truck at the previous time step:

$$0 \le N_{\text{tru,BSS,WB}}^{\omega,n,t} \le C_{\text{tru,cap}} \zeta_{\text{arc},(n,n)}^{\omega,t}, \forall \omega \in H, \forall n \in G_{\text{BSS}}, t \in \mathcal{T} \tag{11}$$

$$\left.\begin{array}{r} N_{\text{tru,BSS,WB}}^{\omega,n,t} = 0, \forall \omega \in H, \forall n \in G_{\text{BSS}}, t = 1 \\ N_{\text{tru,BSS,WB}}^{\omega,n,t} \le S_{\text{tru,WB,stor}}^{\omega,t-1}, \forall \omega \in H, \forall n \in G_{\text{BSS}}, t \in \mathcal{T} \setminus \{1\} \end{array}\right\} \tag{12}$$

Similarly, the loading of WBs and unloading of DBs by trucks at the BCS can be achieved using (13) to (15).

$$0 \le N_{\text{tru,BCS,WB}}^{\omega,m,t} \le C_{\text{tru,cap}} \zeta_{\text{arc},(m,m)}^{\omega,t}, \forall \omega \in H, \forall m \in G_{\text{BCS}}, t \in \mathcal{T} \tag{13}$$

$$0 \le N_{\text{tru,BCS,DB}}^{\omega,m,t} \le C_{\text{tru,cap}} \zeta_{\text{arc},(m,m)}^{\omega,t}, \forall \omega \in H, \forall m \in G_{\text{BSS}}, t \in \mathcal{T} \tag{14}$$

$$\left.\begin{array}{r} N_{\text{tru,BCS,DB}}^{\omega,m,t} = 0, \forall \omega \in H, \forall m \in G_{\text{BCS}}, t = 1 \\ N_{\text{tru,BCS,DB}}^{\omega,m,t} \le S_{\text{tru,DB,stor}}^{\omega,t-1}, \forall \omega \in H, \forall m \in G_{\text{BCS}}, t \in \mathcal{T} \setminus \{1\} \end{array}\right\} \tag{15}$$

where $\zeta_{\text{arc},(m,m)}^{\omega,t}$ is a binary decision variable indicating whether truck $\omega$ is docked at BCS $m$ at time step $t$, if the truck is docked there at that time step, the variable takes the value of 1, if not docked, it takes the value of 0.

*2) Objective Function of the Logistics System Operation Model*

This paper aims to maximize the operational revenue of the logistics system by optimizing the scheduling of trucks and batteries within it, while considering the eVTOL battery swapping demand at BSSs and the battery supply capacity of the BCS. Accordingly, the following objective function for the logistics system operation is presented:

$$\begin{array}{r} \max P_{\text{obj,tran}} = \sum_{t=1}^{T} \Bigg[ -\lambda_{\text{tru,tran}} \sum_{\omega \in H} \beta_{\text{tru_on}}^{\omega,t} \\ -\lambda_{\text{labor}} \sum_{m \in G_{\text{BCS}}} \sum_{\omega \in H} (N_{\text{tru,BCS,DB}}^{\omega,m,t} + N_{\text{tru,BCS,WB}}^{\omega,m,t}) \\ -\lambda_{\text{labor}} \sum_{n \in G_{\text{BSS}}} \sum_{\omega \in H} (N_{\text{tru,BSS,DB}}^{\omega,n,t} + N_{\text{tru,BSS,WB}}^{\omega,n,t}) \Bigg] \end{array} \tag{16}$$

where the first term represents the total transportation cost incurred by the logistics system scheduling, $\lambda_{\text{tru,tran}}$ is the transportation fee for a truck traveling for one time step, the second and third terms are the manual handling costs for loading and unloading batteries at the BCS and BSSs, respectively, and $\lambda_{\text{labor}}$ is the handling cost for a single battery.

*B. BSS Operation Model*

*1) Modeling Principles and Battery Swapping Service Constraints for BSS*

This paper models the operation of the BSS as comprising a queue of eVTOLs awaiting battery swapping, a DB inventory, and a WB inventory. To facilitate the analysis of the BSSs' battery swapping service performance, the discrete cumulative curve method is adopted, modeling the eVTOL queue as the arrival and departure of battery packs[48]. Specifically, the arrival curve is defined as $A^{n,t}$, describing the cumulative number of battery swapping services that BSS $n$ is required to provide from time step 1 to $t$. The departure curve is defined as $D^{n,t}$, recording the cumulative number of battery swapping services that BSS $n$ has actually provided from time step 1 to $t$. Additionally, to prevent passenger waiting times from being unreasonably extended, a minimum departure curve $D_{\min}^{n,t}$ is introduced into the model, which specifies the minimum cumulative number of battery swapping services that BSS $n$ must have completed by the current time step. Assuming there are two CityAirbus aircraft, each with 4 batteries, where CityAirbus1 arrives at BSS $n$ at time step $t$ and can accept departure from the BSS no later than time step $t+2$, and CityAirbus2 arrives at BSS $n$ at time step $t+1$ and can accept departure from the BSS no later than time step $t+3$, then $A^{n,t}$ = 4, $D_{\min}^{n,t+2} = 4$, $A^{n,t+1} = 8$, $D_{\min}^{n,t+3} = 8$.

In summary, the BSS must adhere to the following battery swapping service constraints. First, the $D^{n,t}$ of any BSS at any time step cannot exceed the $A^{n,t}$ of that BSS. Second, to balance the operational revenue of the BCSS and passenger waiting times, the $D^{n,t}$ at each time step must not fall below the specified $D_{\min}^{n,t}$:

$$D_{\min}^{n,t} \le D^{n,t} \le A^{n,t}, n \in G_{\text{BSS}}, t \in \mathcal{T} \tag{17}$$

The second battery swapping service constraint is shown in Equation (18), which serves as a necessary supplement to Equation (17) to ensure that $D^{n,t}$ either increases or remains constant as time steps progress:

$$D^{n,t-1} \le D^{n,t}, n \in G_{\text{BSS}}, t \in \mathcal{T} \setminus \{1\} \tag{18}$$

*2) BSS Battery Dynamic Inventory Model*

During its daily operations, the BSS interacts with the logistics system and eVTOLs for battery exchanges, causing real-time changes in its WB and DB inventories. Accordingly, this section constructs a BSS battery dynamic inventory model as shown in Equations (19)-(25).

$N_{\text{BSS,DB,sup}}^{n,t}$ and $N_{\text{BSS,WB,rec}}^{n,t}$, as defined by Equations (19) and (20), represent the quantity of DBs provided by BSS $n$ to the logistics system and the quantity of WBs received by BSS $n$ from the logistics system at time step $t$, respectively:

$$N_{\text{BSS,DB,sup}}^{n,t} = \sum_{\omega} N_{\text{tru,BSS,DB}}^{\omega,n,t}, \forall \omega \in H, \forall n \in G_{\text{BSS}}, t \in \mathcal{T} \tag{19}$$

$$N_{\text{BSS,WB,rec}}^{n,t} = \sum_{\omega} N_{\text{tru,BSS,WB}}^{\omega,n,t}, \forall \omega \in H, \forall n \in G_{\text{BSS}}, t \in \mathcal{T} \tag{20}$$

Furthermore, the DB and WB inventory of BSS $n$ at time step $t$, namely $S_{\text{BSS,DB,stor}}^{n,t}$ and $S_{\text{BSS,WB,stor}}^{n,t}$, can be defined by the following expressions:

$$S_{\text{BSS,DB,stor}}^{n,t} = S_{\text{BSS,DB,stor}}^{n,\text{ini}} + D^{n,t} - \sum_{h=1}^{t} N_{\text{BSS,DB,sup}}^{n,t}, \forall n \in G_{\text{BSS}}, t \in \mathcal{T} \tag{21}$$

$$S_{\text{BSS,WB,stor}}^{n,t} = S_{\text{BSS,WB,stor}}^{n,\text{ini}} - D^{n,t} + \sum_{h=1}^{t} N_{\text{BSS,WB,rec}}^{n,t}, \forall n \in G_{\text{BSS}}, t \in \mathcal{T} \tag{22}$$

where $S_{\text{BSS,DB,stor}}^{n,\text{ini}}$ and $S_{\text{BSS,WB,stor}}^{n,\text{ini}}$ are the DB and WB inventories of BSS $n$ at the initial time, which can be used to reduce the number of truck trips and waiting times. Specifically, when the battery swap demand of eVTOL is low, BSS $n$ can directly use $S^{n,\mathrm{ini}}_{\mathrm{BSS,WB,stor}}$ to satisfy its swap demand without requesting system dispatch of truck support. After a truck arrives at BSS $n$, if the number of DBs generated by swap services in previous time steps is small, it can directly load DBs from $S^{n,\mathrm{ini}}_{\mathrm{BSS,DB,stor}}$, thereby avoiding long-term truck detention at the BSS.

Similarly, the quantities of batteries loaded, unloaded, and inventoried at BSS $n$ must also be constrained within reasonable ranges. The corresponding constraints are as follows:

$$\left.\begin{array}{l} 0 \le N^{n,t}_{\mathrm{BSS,DB,sup}} \le N_{\mathrm{tru}} C_{\mathrm{tru,cap}}, \forall n \in G_{\mathrm{BSS}}, t \in \mathcal{T} \\ 0 \le N^{n,t}_{\mathrm{BSS,WB,rec}} \le N_{\mathrm{tru}} C_{\mathrm{tru,cap}}, \forall n \in G_{\mathrm{BSS}}, t \in \mathcal{T} \\ 0 \le S^{n,t}_{\mathrm{BSS,DB,stor}} \le 800, \forall n \in G_{\mathrm{BSS}}, t \in \mathcal{T} \\ 0 \le S^{n,t}_{\mathrm{BSS,WB,stor}} \le 800, \forall n \in G_{\mathrm{BSS}}, t \in \mathcal{T} \end{array}\right\} \quad (23)$$

where $N_{\mathrm{tru}}$ is the number of all trucks in the BCSS.

To facilitate the BCSS in initiating operations for the next cycle, at the end of each operational cycle, the WB and DB inventories of BSS $n$ should return to their initial levels. That is, the deviation between the battery inventory at the final time step and the initial inventory must not exceed $\varepsilon$ batteries. This constraint can be described as:

$$\left| S^{n,T}_{\mathrm{BSS,DB,storage}} - S^{n,\mathrm{ini}}_{\mathrm{BSS,DB,storage}} \right| \le \varepsilon, \forall n \in G_{\mathrm{BSS}} \quad (24)$$

$$\left| S^{n,T}_{\mathrm{BSS,WB,storage}} - S^{n,\mathrm{ini}}_{\mathrm{BSS,WB,storage}} \right| \le \varepsilon, \forall n \in G_{\mathrm{BSS}} \quad (25)$$

*3) BSS Objective Function*

The objective of the BSS is to maximize its operational revenue by optimizing the quantity of WBs it receives, the quantity of DBs it provides, and its dynamic inventory. Its objective function is as follows:

$$\max P_{\mathrm{obj,BSS}} = \eta_{\mathrm{batt_swap}} \sum_{n \in G_{\mathrm{BSS}}} D^{n,T} \quad (26)$$

where $\eta_{\mathrm{batt_swap}}$ is the revenue obtained from providing one battery swapping service.

*C. BCS Operation Model*

Building on the integrated consideration of the logistics system model, the BSS operational model, and the operational characteristics of the BCS, this paper constructs an operational model for the BCS. This section elaborates further on core aspects of this model, including the dynamic battery inventory, charging energy, and charging power.

*1) BCS Battery Dynamic Inventory Model*

Equation (27) provides the definition of $N^{m,t}_{\mathrm{BCS,DB,rec}}$, which is the quantity of DBs received by BCS $m$ from the logistics system at time step $t$. It is the sum of DBs unloaded by all trucks parked at this BCS:

$$N^{m,t}_{\mathrm{BCS,DB,rec}} = \sum_{\omega} N^{\omega,m,t}_{\mathrm{tru,BCS,DB}}, \forall \omega \in H, \forall m \in G_{\mathrm{BCS}}, t \in \mathcal{T} \quad (27)$$

Similarly, the quantity of WBs, denoted as $N^{m,t}_{\mathrm{BCS,WB,sup}}$, provided by BCS $m$ to the logistics system can be defined as:

$$N^{m,t}_{\mathrm{BCS,WB,sup}} = \sum_{\omega} N^{\omega,m,t}_{\mathrm{tru,BCS,WB}}, \forall \omega \in H, \forall m \in G_{\mathrm{BCS}}, t \in \mathcal{T} \quad (28)$$

Let $N^{m,t}_{\mathrm{BCS,bin}}$ be the number of DBs placed into charging piles by BCS $m$ at time step $t$, and $N^{m,t}_{\mathrm{BCS,bout}}$ be the number of WBs taken out from charging piles. Therefore, the number of batteries in the charging piles of BCS $m$ at time step $t$ can be expressed as:

$$S^{m,t}_{\mathrm{BCS,in_char}} = \sum_{h=1}^{t} N^{m,h}_{\mathrm{BCS,bin}} - \sum_{h=1}^{t} N^{m,h}_{\mathrm{BCS,bout}}, \forall m \in G_{\mathrm{BCS}}, t \in \mathcal{T} \quad (29)$$

Let $N^{m,t}_{\mathrm{BCS,back}}$ first denote the quantity of backup DBs deployed by BCS $m$ at time step $t$. Combining Equations (27), (28), and (29), the DB inventory $S^{m,t}_{\mathrm{BCS,DB,stor}}$ and WB inventory $S^{m,t}_{\mathrm{BCS,WB,stor}}$ of the BCS can be derived as follows:

$$\begin{aligned} S^{m,t}_{\mathrm{BCS,DB,stor}} = \sum_{h=1}^{t} (N^{m,h}_{\mathrm{BCS,back}} + N^{m,h}_{\mathrm{BCS,DB,rec}} \\ - N^{m,h}_{\mathrm{BCS,bin}}), \forall m \in G_{\mathrm{BCS}}, t \in \mathcal{T} \end{aligned} \quad (30)$$

$$S^{m,t}_{\mathrm{BCS,WB,stor}} = \sum_{h=1}^{t} (N^{m,h}_{\mathrm{BCS,bout}} - N^{m,h}_{\mathrm{BCS,WB,sup}}), \forall m \in G_{\mathrm{BCS}}, t \in \mathcal{T} \quad (31)$$

To ensure the completeness of the battery charging process, a minimum charging time constraint must be introduced. Constraint (32) prevents the charging process from being prematurely interrupted, thereby ensuring that each battery can reach a usable state of charge.

$$\sum_{h=1}^{t} N_{\mathrm{BCS,bout}}^{m,h} \le \sum_{h=1}^{t-(T_{\mathrm{C}}-1)} N_{\mathrm{BCS,bin}}^{m,h}, \forall m \in G_{\mathrm{BCS}}, t \in \mathcal{T} \tag{32}$$

where $T_{\mathrm{C}}$ is the minimum time required to fully charge a battery.

All decision variables of the BCS must also remain within reasonable bounds. Their expressions can be formulated as:

$$\left.\begin{array}{c}
0 \le N_{\mathrm{BCS,bin}}^{m,t} \le N_{\mathrm{charge}}, \forall m \in G_{\mathrm{BCS}}, t \in \mathcal{T} \\
0 \le N_{\mathrm{BCS,bout}}^{m,t} \le N_{\mathrm{charge}}, \forall m \in G_{\mathrm{BCS}}, t \in \mathcal{T} \\
0 \le N_{\mathrm{BCS,back}}^{m,t} \le 1000, \forall m \in G_{\mathrm{BCS}}, t \in \mathcal{T} \\
0 \le N_{\mathrm{BCS,DB,rec}}^{m,t} \le N_{\mathrm{tru}} C_{\mathrm{tru,cap}}, \forall m \in G_{\mathrm{BCS}}, t \in \mathcal{T} \\
0 \le N_{\mathrm{BCS,WB,sup}}^{m,t} \le N_{\mathrm{tru}} C_{\mathrm{tru,cap}}, \forall m \in G_{\mathrm{BCS}}, t \in \mathcal{T} \\
0 \le S_{\mathrm{BCS,in_char}}^{m,t} \le N_{\mathrm{charge}}, \forall m \in G_{\mathrm{BCS}}, t \in \mathcal{T} \\
0 \le S_{\mathrm{BCS,DB,stor}}^{m,t} \le 1000, \forall m \in G_{\mathrm{BCS}}, t \in \mathcal{T} \\
0 \le S_{\mathrm{BCS,WB,stor}}^{m,t} \le 1000, \forall m \in G_{\mathrm{BCS}}, t \in \mathcal{T}
\end{array}\right\} \tag{33}$$

where $N_{\mathrm{charge}}$ is the number of charging piles at the BCS, set to 300 charging piles.

Furthermore, certain restrictions must be imposed on the time steps and quantities concerning the WBs taken out from the charging piles by the BCS and the WBs supplied to the logistics system. Equation (34) stipulates that WBs can only be taken out from the charging piles starting from time step 2, and the quantity taken out cannot exceed the total number of batteries in the charging piles at the previous time step.

$$\left.\begin{array}{c}
N_{\mathrm{BCS,bout}}^{m,t} = 0, \forall m \in G_{\mathrm{BCS}}, t = 1 \\
N_{\mathrm{BCS,bout}}^{m,t} \le S_{\mathrm{BCS,in_char}}^{m,t-1}, \forall m \in G_{\mathrm{BCS}}, t \in \mathcal{T} \setminus \{1\}
\end{array}\right\} \tag{34}$$

Equation (35) stipulates that the BCS cannot supply WBs to the logistics system at time step 1. At other time steps, it may supply WBs, but the quantity must not exceed the WB inventory of the BCS at the previous time step.

$$\left.\begin{array}{c}
N_{\mathrm{BCS,WB,supply}}^{m,t} = 0, \forall m \in G_{\mathrm{BCS}}, t = 1 \\
N_{\mathrm{BCS,WB,supply}}^{m,t} \le S_{\mathrm{BCS,WB,storage}}^{m,t-1}, \forall m \in G_{\mathrm{BCS}}, t \in \mathcal{T} \setminus \{1\}
\end{array}\right\} \tag{35}$$

*2) Energy and Power Constraints of the BCS*

The operation of the BCS must consider the cost of electricity procurement. To control this cost, it must be ensured that all energy consumed by BCS $m$ is used for battery charging. The corresponding expression is as follows:

$$\sum_{h=1}^{t} (P_{\mathrm{BCS,char}}^{m,h} \alpha_{\mathrm{effi,char}} \Delta t) = E_{\mathrm{batt,en}} \sum_{h=1}^{t} N_{\mathrm{BCS,bout}}^{m,h}, \forall m \in G_{\mathrm{BCS}}, t \in \mathcal{T} \tag{36}$$

where $P_{\mathrm{BCS,char}}^{m,t}$ is the total charging power supplied by BCS $m$ at time step $t$, which is the sum of the charging power required by all batteries being charged, $\alpha_{\mathrm{effi,char}}$ is the charging efficiency, $E_{\mathrm{batt,en}}$ is the single battery capacity.

For $P_{\mathrm{BCS,char}}^{m,t}$ in Equation (36), Equation (37) defines its adjustable range during optimization:

$$0 \le P_{\mathrm{BCS,char}}^{m,t} \le P_{\mathrm{max,char}} N_{\mathrm{charge}} \beta_{\mathrm{BCS,char,sign}}^{m,t}, \forall m \in G_{\mathrm{BCS}}, t \in \mathcal{T} \tag{37}$$

where $P_{\mathrm{max,char}}$ is the maximum charging power that each charging pile can supply to a battery, $\beta_{\mathrm{BCS,char,sign}}^{m,t}$ is a binary variable indicating the charging status of BCS $m$, taking the value 1 if the BCS is permitted to charge, and 0 otherwise.

*3) BCS Objective Function*

The primary objective of the BCS is to meet the WB demand of BSSs while adhering to all operational constraints. Under this premise, to maximize operational revenue, the BCS must formulate an optimal operational strategy by optimizing the timing of electricity procurement, the magnitude of charging power, and the quantity of backup DBs used. Its objective function is as follows:

$$\begin{aligned}
\max P_{\mathrm{obj,BCS}} = \sum_{m \in G_{\mathrm{BCS}}} \sum_{t=1}^{T} \big[ & -(\lambda_{\mathrm{pri,mark}}^{t} + \lambda_{\mathrm{pri,batt}}) P_{\mathrm{BCS,char}}^{m,t} \\
& -\lambda_{\mathrm{pri,back}} N_{\mathrm{BCS,back}}^{m,t} \big], \forall t \in \mathcal{T}
\end{aligned} \tag{38}$$

where the first term covers two operational costs of the BCS over $T$ time steps: the first is the total electricity purchase cost, and the second is the total battery degradation compensation cost incurred by the charging process, the second term is the total cost incurred from deploying backup DBs, which refers to the fees paid by the BCS to a third-party company for leasing batteries when it urgently needs to meet the BSS's WB demand but finds its own DB inventory insufficient. $\lambda_{\text{pri,mark}}^{t}$ is the electricity procurement unit price at time step $t$ ; $\lambda_{\text{pri,batt}}$ is the unit price for compensating battery degradation; $\lambda_{\text{pri,back}}$ is the cost of deploying a backup DB.

Integrating all the models above, this paper establishes the following total objective function and operational constraints for the BCSS to achieve optimal system scheduling. Since the research problem constitutes a mixed-integer linear programming problem, the commercial solver Gurobi is employed to solve it using a centralized method.

$$\left.\begin{array}{l}\max P_{\text{obj,all}} = P_{\text{obj,tran}} + P_{\text{obj,BSS}} + P_{\text{obj,BCS}} \\ \text{s.t.}\ \ (1)\text{-}(15),(17)\text{-}(25),(27)\text{-}(37).\end{array}\right\} \quad (39)$$

## V. CASE STUDY

To validate the effectiveness of the eVTOL battery charging and swapping system model based on CLSC and TSN proposed in this paper, this section conducts a simulation analysis for the scenario illustrated in Fig. 1. The case study is programmed using MATLAB R2022b, and the commercial solver Gurobi is called for the solution.

In the simulation, the BCSS managed by the eVTOL operator consists of several eVTOLs requiring battery swapping, four trucks, three BSSs, and one BCS. The operational cycle $T$ is set to twenty-four time steps, with the objective of maximizing the comprehensive operational revenue of the entire system. All four trucks are located at the BCS at the initial time step. The $A^{n,t}$ and $D_{\min}^{n,t}$ for each BSS at each time step are shown in Fig.4. The electricity procurement prices during the operational cycle are sourced from data published by Singapore's EMC [49] on December 15, 2017, as shown in Fig. 5. The electricity price fluctuates frequently and with drastic amplitude. If the BCSS proposed in this paper can effectively respond to this characteristic during operation and formulate a reasonable operational plan, it will strongly demonstrate that the system possesses excellent flexibility and robustness. The remaining simulation parameters are summarized in Table I. The solution time and key operational results of the BCSS are presented in Table II. To thoroughly evaluate the system performance, the simulation results of the logistics system operational model, the BSS operational model, and the BCS operational model will be analyzed and discussed separately. Finally, a sensitivity analysis of the model is conducted to demonstrate its excellent performance in multiple aspects, including battery swap service quality, system sustainability, and operational resilience.

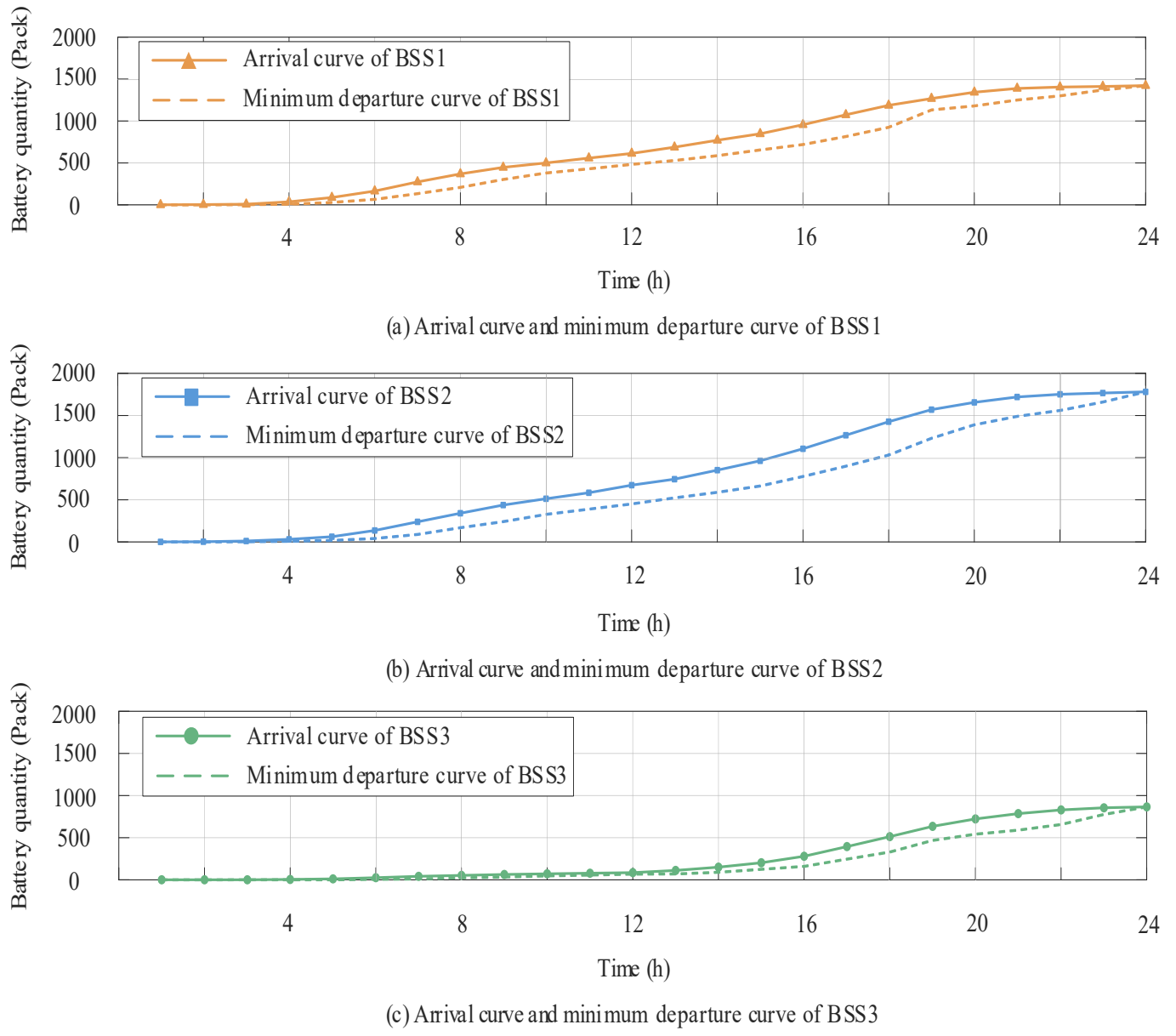

Fig. 4. Arrival curve and minimum departure curve of BSSs.

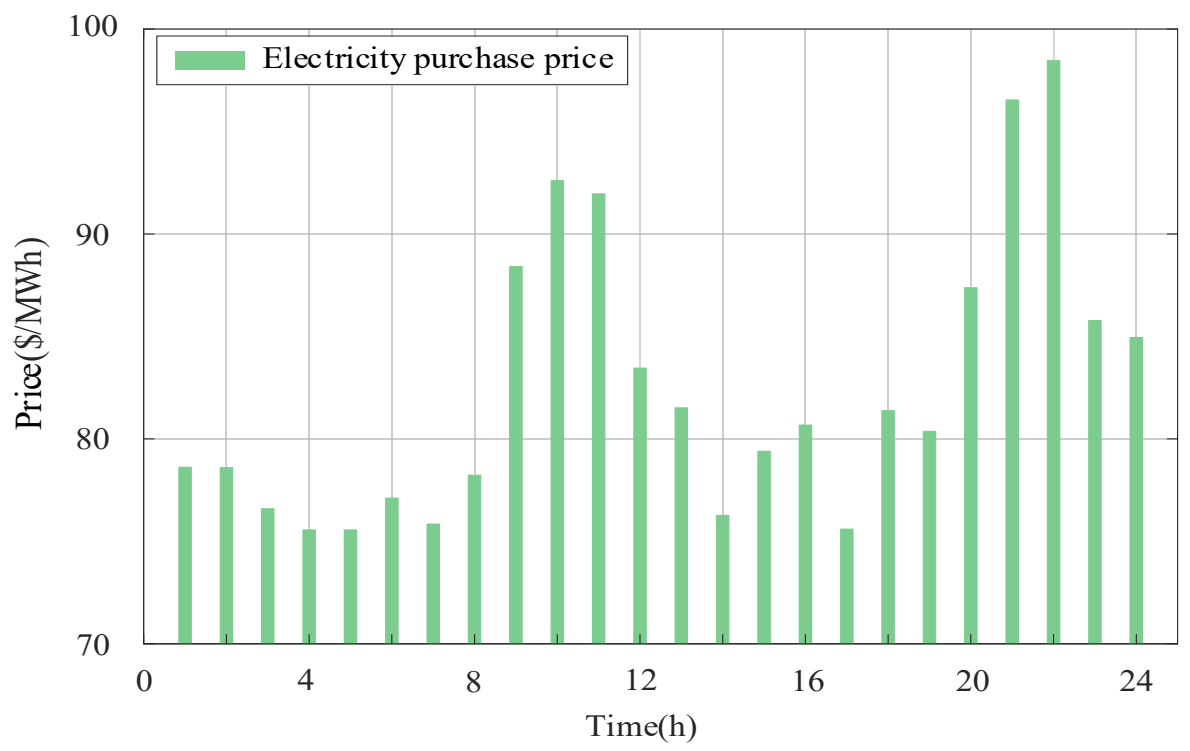

Fig. 5. Electricity purchase price of BCS.

TABLE I. PARAMETERS OF BCSS

| ***Parameter*** | ***Value*** | ***Parameter*** | ***Value*** |
|---|---|---|---|
| $C_{\text{tru,cap}}$ (pack) | 300 | $\lambda_{\text{pri,back}}$ ($/pack) | 5 |
| $\lambda_{\text{tru,tran}}$ ($/h) | 10 | $\lambda_{\text{pri,batt}}$ ($/kWh) | 0.01 |

| ***Parameter*** | ***Value*** | ***Parameter*** | ***Value*** |
|---|---|---|---|
| $\lambda_{\text{labor}}$ ($/pack) | 0.1 | $E_{\text{batt,en}}$ (kWh) | 30 |
| $S_{\text{BSS,DB,stor}}^{n,\text{ini}}$ (pack) | 200 | $\alpha_{\text{effi,char}}$ | 0.95 |
| $S_{\text{BSS,WB,stor}}^{n,\text{ini}}$ (pack) | 200 | $\varepsilon$ (pack) | 30 |
| $\eta_{\text{batt_swap}}$ ($/pack) | 6 | $P_{\text{max,char}}$ (kW) | 120 |

TABLE II. OPERATIONAL RESULTS OF BCSS

| $P_{\text{obj,tran}}$ | $P_{\text{obj,BSS}}$ | $P_{\text{obj,BCS}}$ | ***Solution Time*** |
|---|---|---|---|
| -2021.6 $ | 24414 $ | -16059.82 $ | 10129.79 s |

### *A. Feasibility Analysis of the BCSS*

#### *1) Analysis of Logistics System Simulation Results*

Fig. 6 to 9 present the scheduling results of the four trucks over the operational cycle, verifying the feasibility of the proposed logistics system model. As shown in Fig. 6(a), 7(a), and 9(a), trucks1, 2, and 4 are scheduled by the model to travel to different BSSs at the early stage of operations to bring back a large number of DBs as soon as possible, thereby significantly reducing the BCS's usage of spare DBs and operational costs. The scheduling of truck3 is completely opposite. As shown in Fig. 8(b), this truck waits at the BCS until time step4 to finish loading WBs, and then delivers them to BSS1 at time step5. From Fig. 4, it can be seen that BSS1 has the highest battery swap demand among all BSSs before time step8. Without the timely replenishment by truck3, its WB inventory at the beginning of time step 7 would be insufficient to meet the swap demand within that time step. Observing Fig. 6 and 7, it can also be found that truck 1 does not immediately unload DBs at time step4; instead, it first loads 79 WBs, and then delivers 201 DBs to the BCS at time step 5. Truck2, at time step 14, first unloads 300 DBs at the BCS and then loads 300 WBs. This indicates that the logistics system model can intelligently adjust the loading and unloading sequence based on the actual status of the trucks.

From Fig. 4, it can be seen that the battery swap demands of BSS1 and BSS2 are both significantly higher than that of BSS3. Analyzing the travel trajectories of the four trucks reveals that battery swap demand plays a decisive role in route planning and service allocation. Truck2 primarily serves BSS2, only occasionally stopping at BSS1 and BSS3 to load and unload batteries when passing by. Truck4 focuses on serving BSS1, shuttling between the BCS and BSS1 throughout time steps 1 to 20. The trajectory of Truck1 is relatively more complex, mainly moving among BSS2, BSS3, and the BCS, and only stopping at BSS1 at the final time step to load and unload batteries. Truck3 exhibits the most balanced dispatch, covering all stations in its service range. It can be seen that the WB demand of BSS3 is jointly satisfied by Truck1 and Truck3. In summary, the scheduling of the logistics system model is both intelligent and economical. On the one hand, it dynamically plans routes based on battery swap demand and flexibly adjusts loading and unloading strategies according to the status of the trucks. On the other hand, by arranging for some trucks to bring back DBs as early as possible to reduce operational costs while keeping other trucks on standby to respond quickly to BSS demand, it achieves a synergistic optimization between resource allocation and cost control.

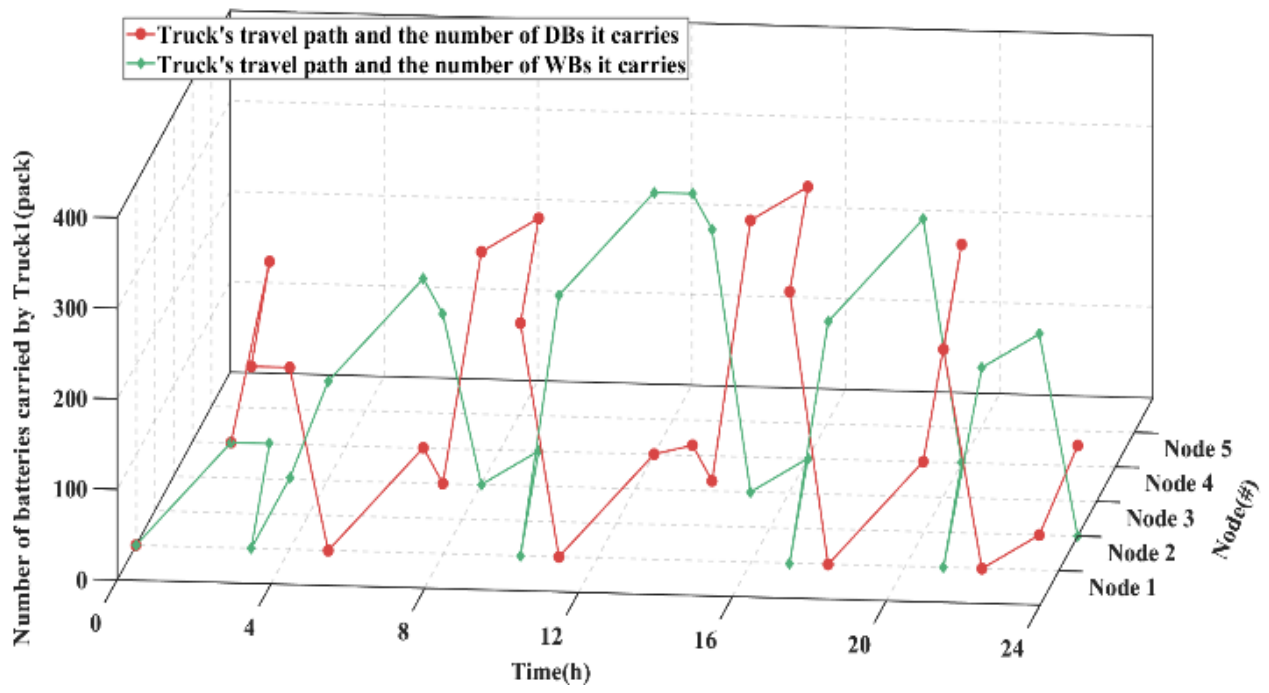

Fig. 6. Driving trajectory and change in battery inventory of Truck1.

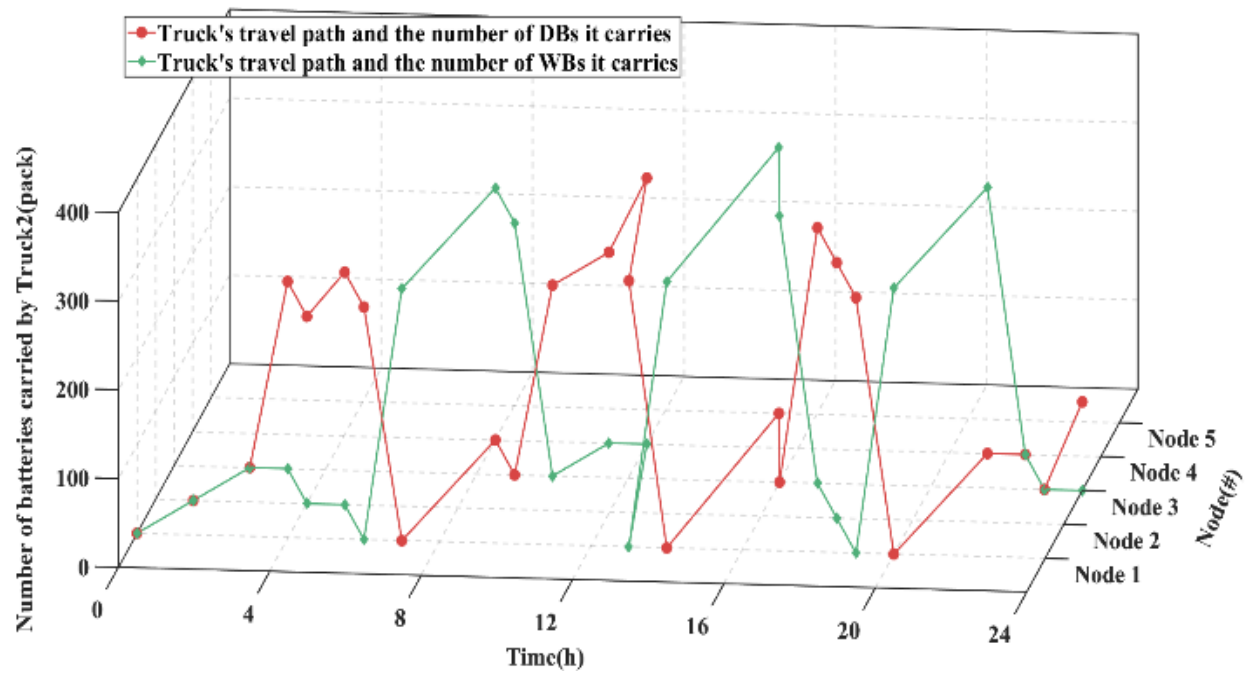

Fig. 7. Driving trajectory and change in battery inventory of Truck2.

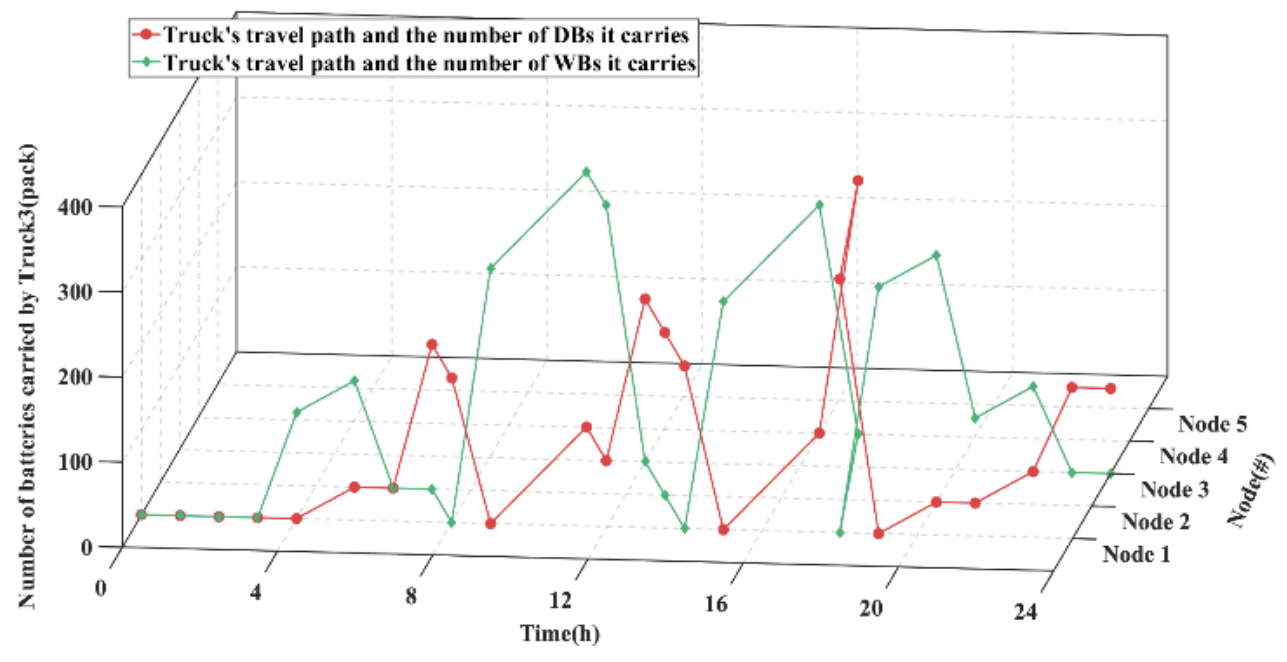

Fig. 8. Driving trajectory and change in battery inventory of Truck3.

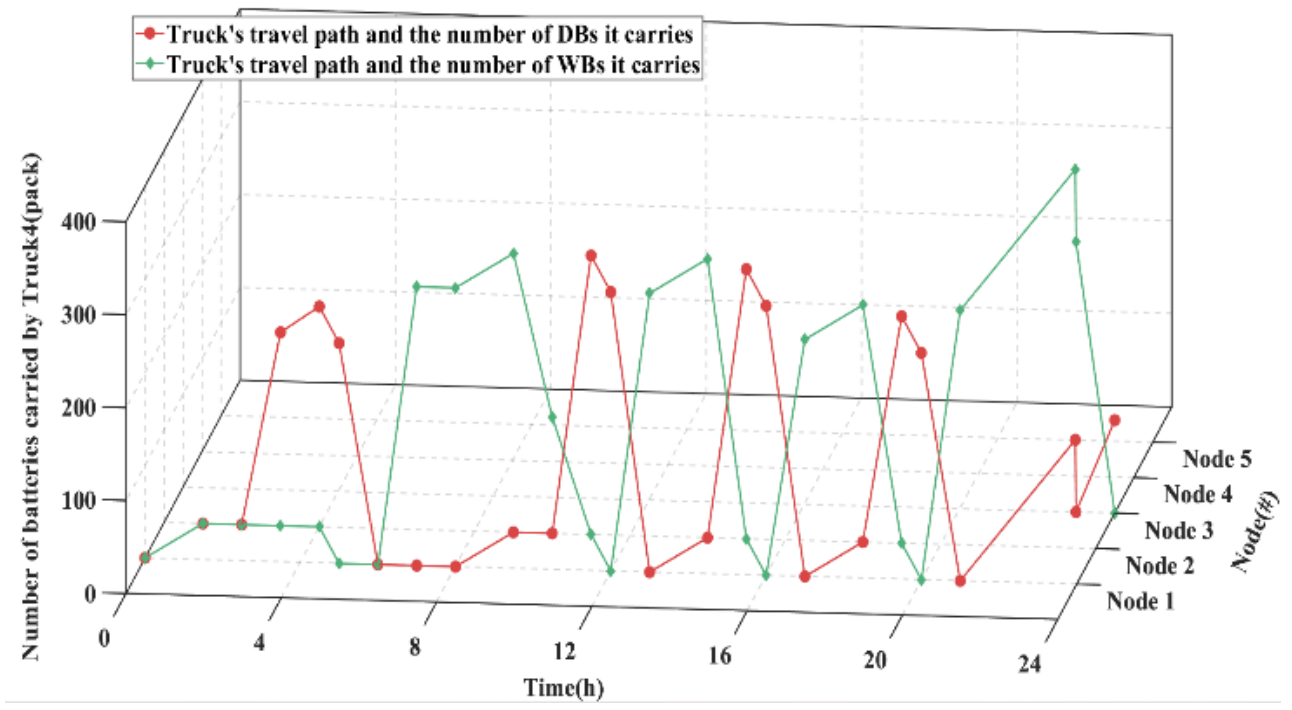

Fig. 9. Driving trajectory and change in battery inventory of Truck4.

*2) Analysis of BSS Simulation Results*

From Fig. 4 and 10, it can be seen that each BSS experiences low battery swap pressure during time steps1 to 6, and thanks to the initial WB inventories with which they are equipped, the $D^{n,t}$ of each station remains essentially consistent with its $A^{n,t}$. As operations progress, when a certain BSS reaches a critical state where $D^{n,t}$ equals $D^{n,t}_{\min}$, the BCSS promptly arranges for trucks to deliver a large number of WBs in subsequent time steps. Taking BSS1 as an example, at the end of time step9, its $D^{n,t}$ has become equal to its $D^{n,t}_{\min}$. Subsequently, Truck4 unloads 300 WBs in batches to this station at time steps10 and 11. Nevertheless, BSS1 still faces significant operational pressure, even exhibiting a phenomenon at some time steps where it "has a small amount of WB inventory but does not provide battery swap services." Analyzing Fig. 4, 10, and 11, it can be seen that BSS1 has a limited supply of WBs but high swap pressure. Therefore, before providing swap services, it must carefully plan and utilize its $S^{n,t}_{\mathrm{BSS,WB,stor}}$, while also reserving some inventory for subsequent operations as much as possible, subject to the constraint that $D^{n,t}$ cannot be less than $D^{n,t}_{\min}$. For example, at time steps13 and 14, BSS1 receives no WBs from trucks and has only 11 WBs remaining in inventory. However, its $D^{n,t}$ has reached 614 by this time, fully satisfying the battery swap service constraint. Even if it uses up all WBs during these two time steps, it would still be unable to achieve consistency between $D^{n,t}$ and $A^{n,t}$, resulting in a situation of "having inventory but zero swap services." After receiving 300 WBs at time step15, BSS1 is able to satisfy all previous swap demands at once. Similar situations occur at time steps16 and 17, but the key difference is that no truck delivers WBs at time step18, so BSS1 can only use up its remaining 77 WBs, barely ensuring that $D^{n,t}$ is not less than $D^{n,t}_{\min}$. An in-depth analysis reveals that the low frequency of truck service and the small number of WBs provided by the BCSS to BSS1 are the root causes of the above phenomenon. The BCSS needs to reduce battery charging costs and truck scheduling costs to improve overall operational revenue. However, the means by which the BSS ensures the quality of battery swap services is to perform swaps for eVTOL as quickly as possible, creating a conflict between external battery supply and internal demand during BSS operations. Nevertheless, based on the above discussion, it is not difficult to see that the BSS proposed in this paper achieves an effective balance between the two and formulates a reasonable battery swap plan.

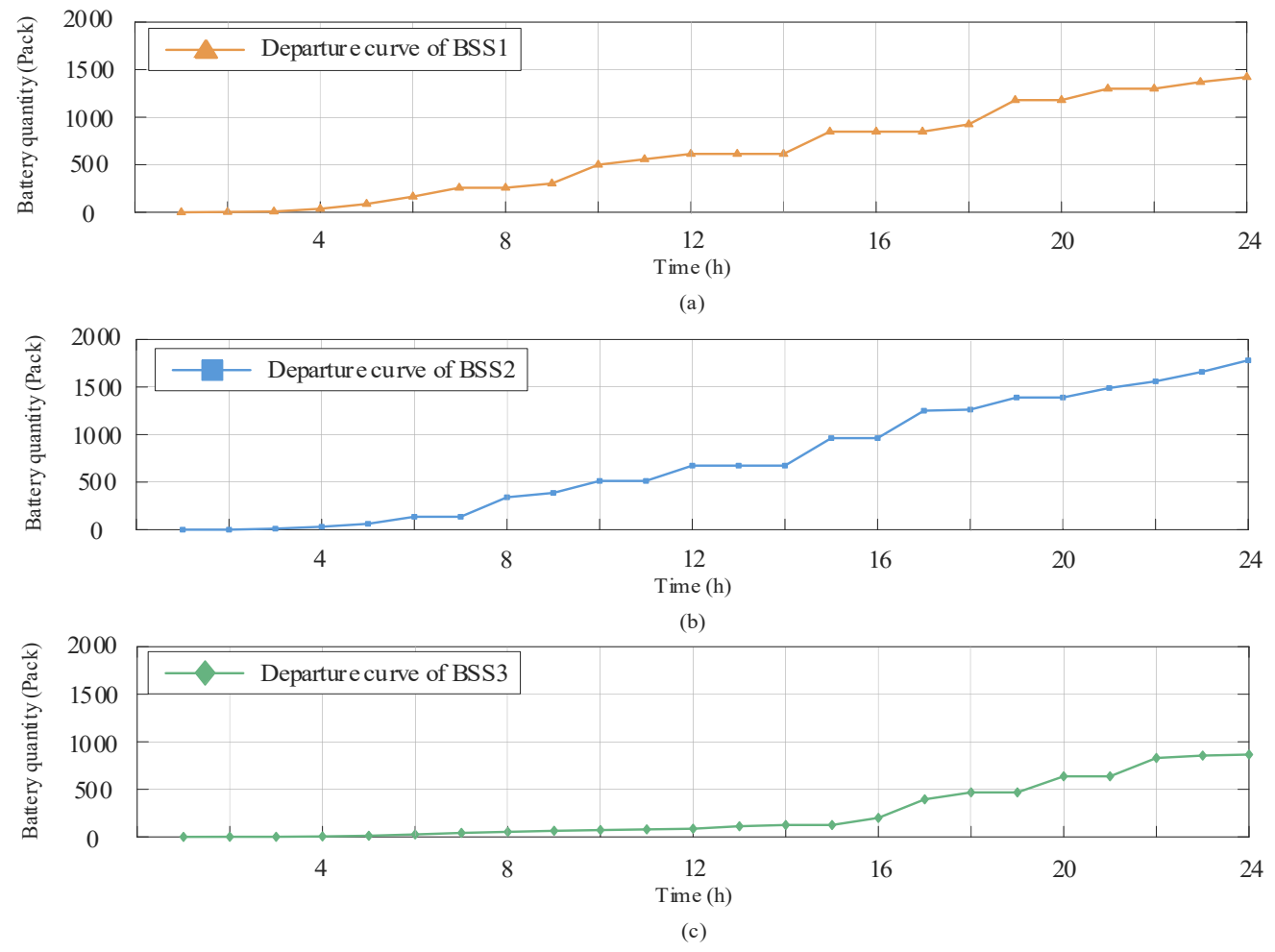

Fig. 10. Battery-swap result of the BSS Model.

From Fig. 12, it can be seen that at time steps 7, 11, 13, and 14, BSS2 also experiences delayed battery swapping while still having WB inventory remaining. However, its $D^{n,t}$ performs significantly better than that of BSS1, stabilizing in the middle of the interval formed by $A^{n,t}$ and $D_{\min}^{n,t}$. Overall, although BSS2 bears the highest eVTOL battery swap demand among all stations, it receives sufficient WB replenishment during most of the operational period, and thus its operational status during time steps1 to 19 is relatively relaxed.

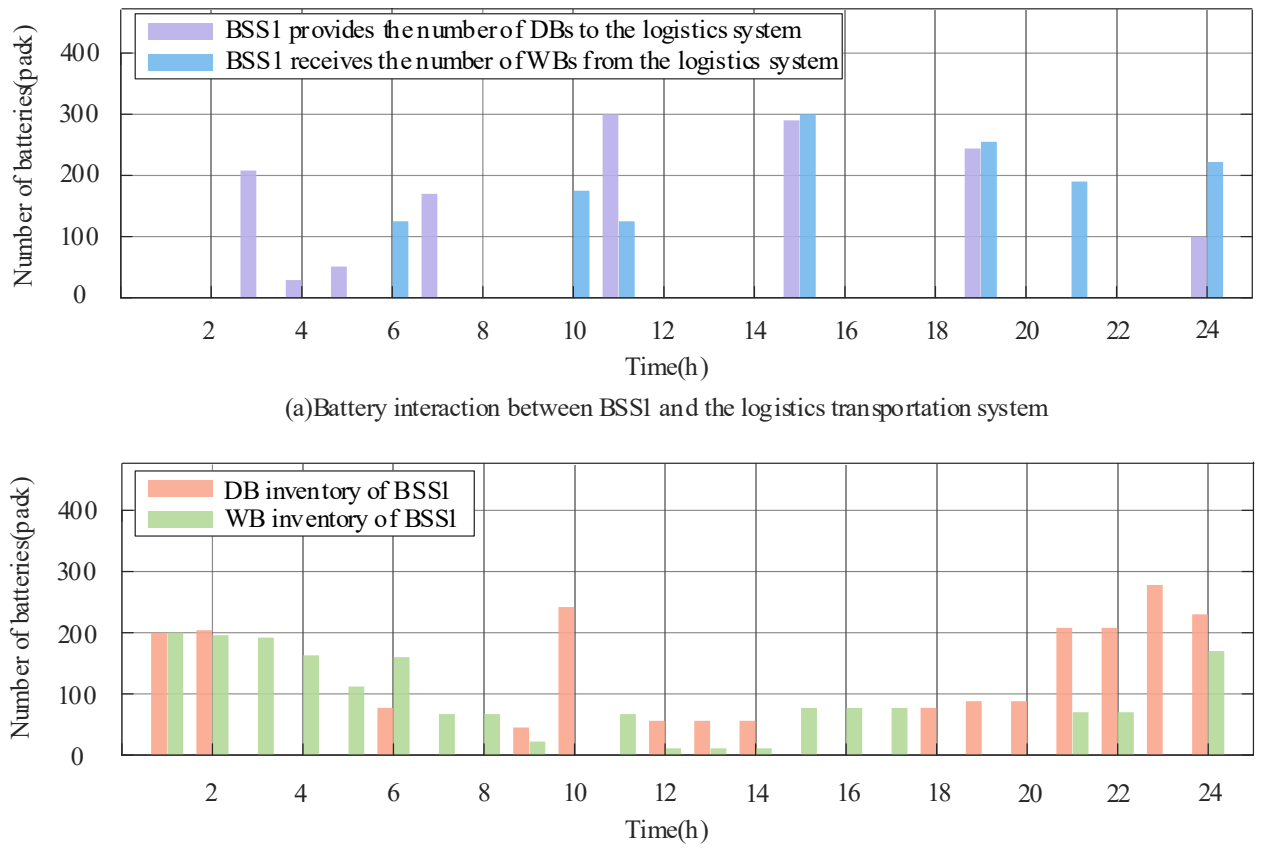

(a)Battery interaction between BSS1 and the logistics transportation system

(b)Battery inventory of BSS1

Fig. 11. Operational results of BSS1.

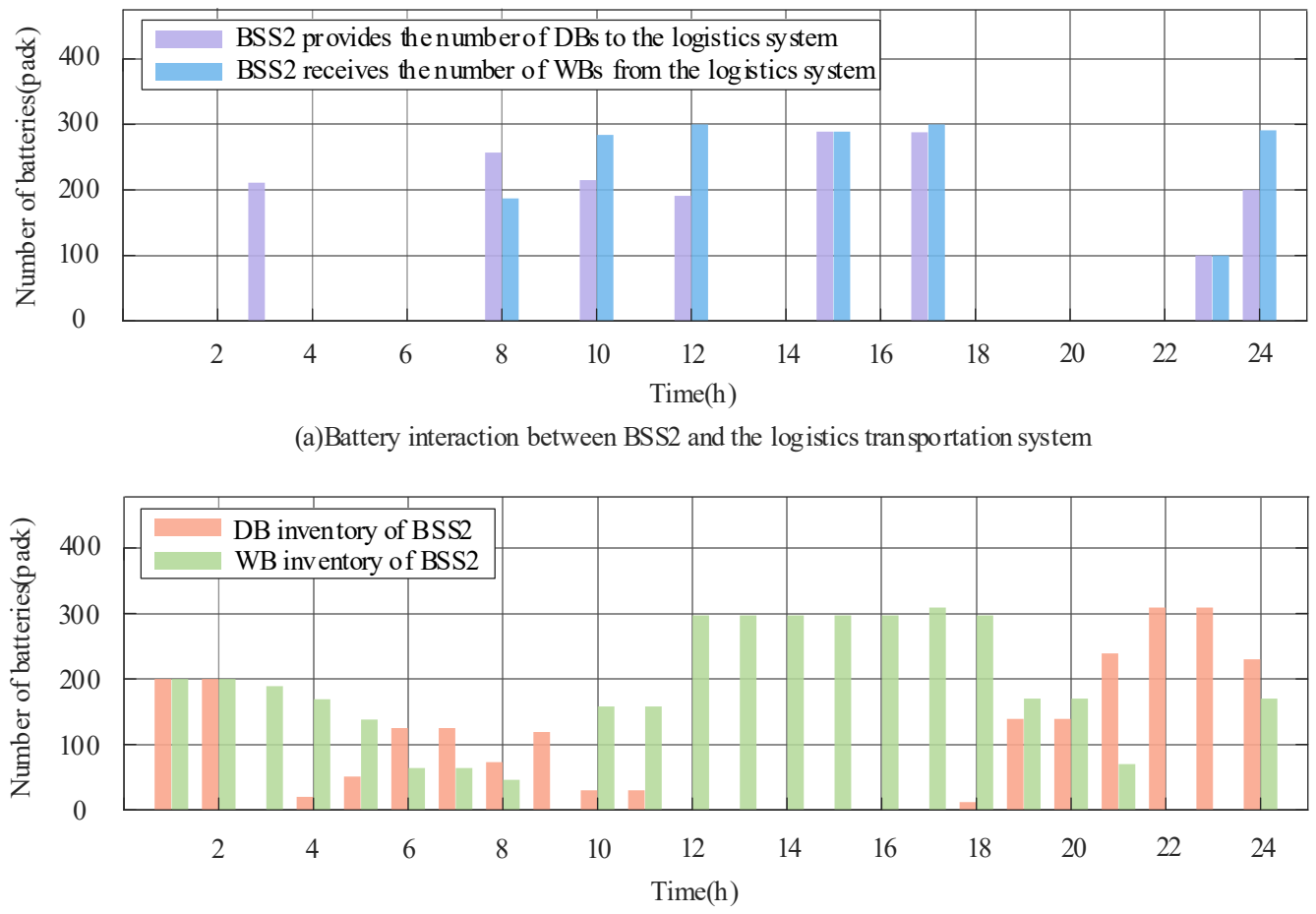

(a)Battery interaction between BSS2 and the logistics transportation system

(b)Battery inventory of BSS2

Fig. 12. Operational results of BSS2.

According to the operational status shown in Fig. 13, BSS3 receives no WBs from trucks during the first 13 time steps. However, due to its low swap demand, it is able to maintain a high quality of battery swap service relying solely on its initial WB inventory. Over the entire operational cycle, BSS3 provides DBs to the logistics system only five times, which is significantly lower than the other BSSs in terms of both delivery frequency and delivery quantity.

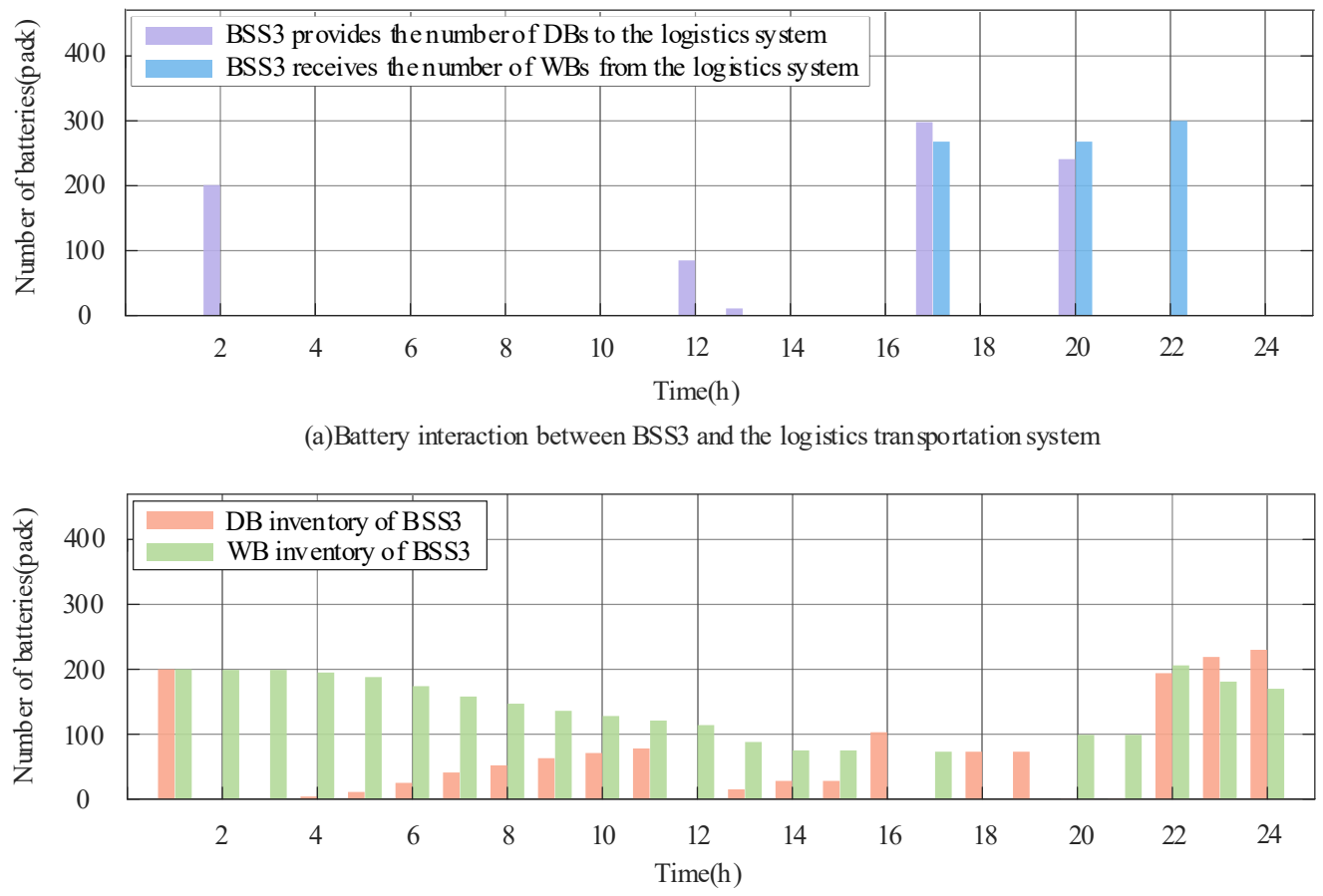

(a)Battery interaction between BSS3 and the logistics transportation system

(b)Battery inventory of BSS3

Fig. 13. Operational results of BSS3.

### 3) *Analysis of BCS Simulation Results*

As shown in Fig. 14, thanks to the assistance provided by Trucks1, 2, and 4 at the early stage of operations, the BCS only needs to call upon spare DBs four times during time steps2 to 6, accumulating a total of 958 DBs, which significantly reduces battery leasing costs. From Fig. 5 and 14, it can be seen that because electricity prices are relatively high during time steps9 to 11, the BCS does not place DBs into chargers at time steps9 and 10; even when 300 DBs are placed into chargers at time step11, charging does not begin immediately. From Fig. 15, it can be observed that the BCS charges only 127 DBs at time step12, removes the WBs from the chargers after they are fully charged, and then promptly places 127 DBs back into the chargers. This phenomenon indicates that the DBs placed into chargers at time step11 do not represent a "bottleneck" in the model, but rather a normal scheduling behavior driven by the BSS's demand for WBs in subsequent time steps. The reason only 127 DBs are charged is that, although the electricity price at time step12 has decreased compared to time steps9 to 11, it remains higher than the prices during time steps13 to 19. To timely satisfy BSS1's demand for WBs while minimizing charging costs as much as possible, the BCS must place DBs into chargers in advance and consequently charge only a portion of them. As described above, BSS1 delays battery swapping at time steps13 and 14, has only 11 WBs in inventory, and receives no WB deliveries from trucks. Therefore, the WBs removed from chargers at time step12 are precisely those used to support BSS1 at time step15. Specifically, the BCS has already accumulated 173 WBs at time step11, reaching 300 WBs by time step12. Truck4 loads these WBs at time step13 and transfers them to BSS1 at time step15. Entering time steps20 to 21, electricity prices once again reach peak levels. However, to satisfy the demands of each BSS, the BCS must arrange charging during this period. To reduce charging costs, the BCS again charges only a portion of the batteries in the chargers at time step21. During time steps22 to 24, the number of batteries in the chargers remains at 300, while the BCS's WB inventory remains consistently at zero. The reason for this phenomenon is that electricity prices during this stage are too high, and the WBs previously accumulated have already been sufficient to meet the battery swap demands of the BSSs, so no additional charging operations are performed.

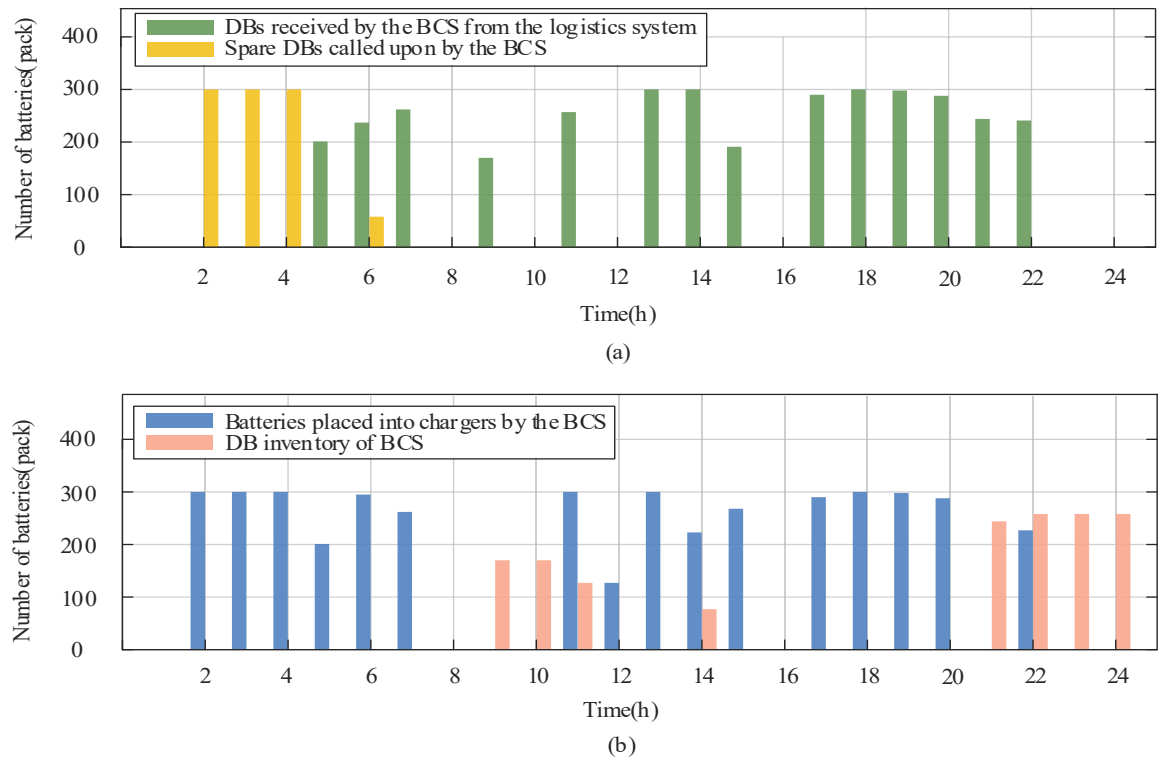

Fig. 14. BCS's DB inventory and its related variables.

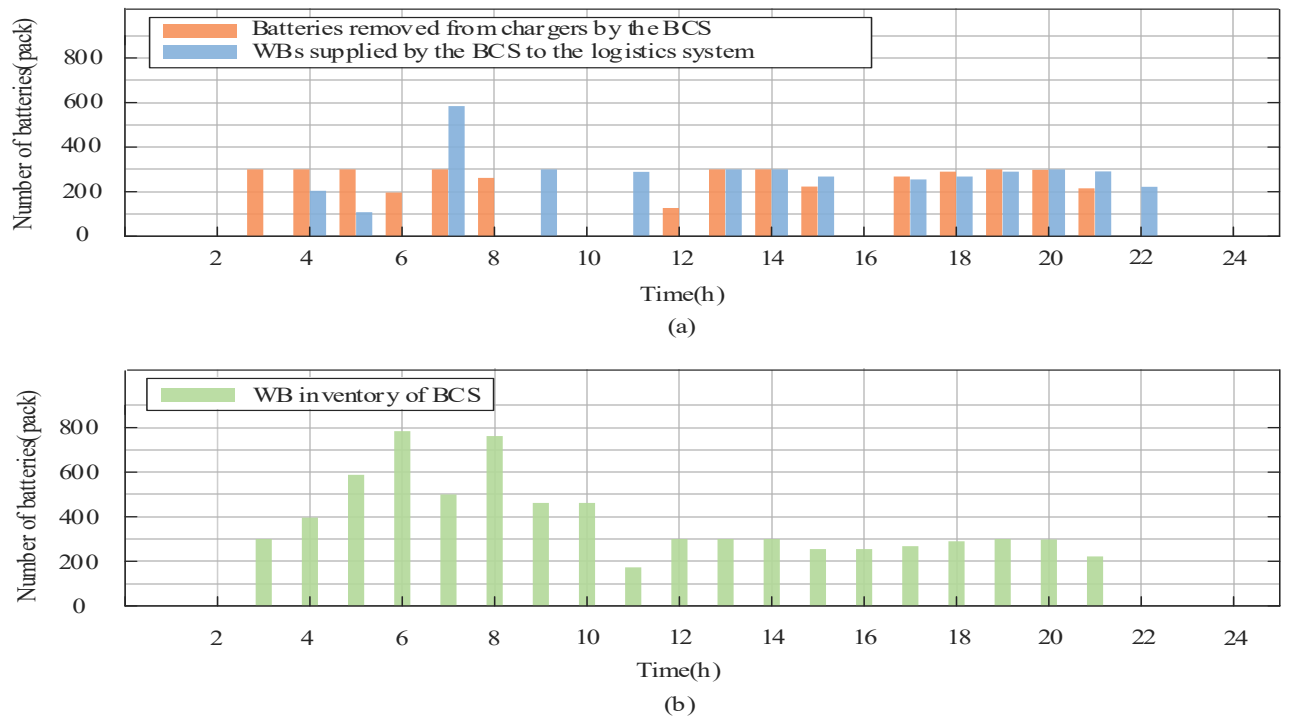

Fig. 15. BCS's WB inventory and its related variables.

From the above analysis, it can be seen that the BCS constructed in this paper performs well in simultaneously accommodating the WB demand of the BSSs and formulating a reasonable charging schedule. Furthermore, the above operational process integrates the three key links of charging scheduling, transportation and distribution, and battery swapping, clearly demonstrating the efficient coordination among the members of the BCSS in both time and space dimensions.

### B. *Sensitivity Analysis*

#### 1) *Comparative Analysis Between the Proposed BCSS and a Simplified Benchmark Model That Does Not Consider Battery Swap Service Quality*

When formulating a BCSS scheduling scheme, if operators focus solely on maximizing operational revenue, it is very easy to neglect battery swap service quality. Under such a scheduling strategy, passenger waiting time at BSSs may be unreasonably prolonged. However, from the perspectives of commercialization and sustainable development, customer recognition of service quality is the foundation for broad market acceptance. Once the battery swapping link becomes an operational bottleneck, even if the revenue is temporarily optimal, long-term operation becomes difficult to sustain. Therefore, this section conducts a multi-dimensional comparison between the proposed BCSS and a simplified benchmark model that does not consider service quality, to fully validate the superiority of the proposed BCSS in terms of battery swap service quality.

On the surface, the total operational revenue of the simplified benchmark model in Table III is USD 1,624.57 higher than that of the proposed BCSS, and its operational costs for both the BCS and the logistics system are also lower. However, this economic "advantage" does not imply that the model is inherently superior. An in-depth analysis of the operational data of this model reveals that the numbers of spare DBs called upon, DBs received, and WBs supplied by its BCS over the operational cycle are 45, 2,922, and 2,967, respectively, whereas the corresponding figures for the proposed BCSS reach 958, 3,579, and 3,979. The significant differences in these key indicators directly result in lower battery leasing costs, charging costs, and battery loading and unloading costs for the simplified benchmark model compared to the proposed BCSS. Further examining the operational status of the logistics system reveals that the four trucks in this model are in a docked state for a total of 56 time steps over the operational cycle, compared to only 49 time steps in the proposed BCSS. Thus, lower truck utilization also contributes to the lower logistics costs. The core reason why this simplified model exhibits a "false advantage" is that it deviates from the fundamental function of the BCSS, i.e., to provide sufficient WBs for all eVTOLs and ensure the quality of battery swap services, a point that can be further verified by the $A^{n,t}$ and $D^{n,t}$ values of each BSS.

TABLE III. OPERATIONAL RESULTS OF THE SIMPLIFIED BENCHMARK MODEL

| $P_{obj,tran}$ | $P_{obj,BSS}$ | $P_{obj,BCS}$ | $P_{obj,all}$ |
|---|---|---|---|
| -1582.3 \$ | 18342 \$ | -8802.55 \$ | 7957.15 \$ |

From Fig. 16, it can be seen that the $D^{n,t}$ of BSS2 in the simplified benchmark model remains constant at 745 during time steps14 to 23, indicating that this BSS provides no battery swap services to any eVTOL for ten consecutive time steps. The inability of the BSS to supply battery replenishment to eVTOL within a reasonable timeframe directly leads to a severe decline in the operational efficiency of the eVTOL fleet, introducing intolerable deviations into the simulation results. Furthermore, the $A^{n,t}$ corresponding to BSS1, BSS2, and BSS3 are 1,422, 1,781, and 866, respectively, while the $D^{n,t}$ at the end of operations for this model are only 1,304, 1,020, and 733. This comparison reveals that, in its pursuit of revenue maximization, the model fails to complete the task of providing services to all eVTOL awaiting battery swapping, ultimately forcing a large number of passengers to remain stranded at BSSs for extended periods. It should be noted that this simplified benchmark model does not account for costs that are unavoidable in real-world operations, such as ticket refunds, delay compensation, and complaint penalties resulting from passenger stranding. Otherwise, its overall operational revenue would certainly be far lower than that of the proposed BCSS.

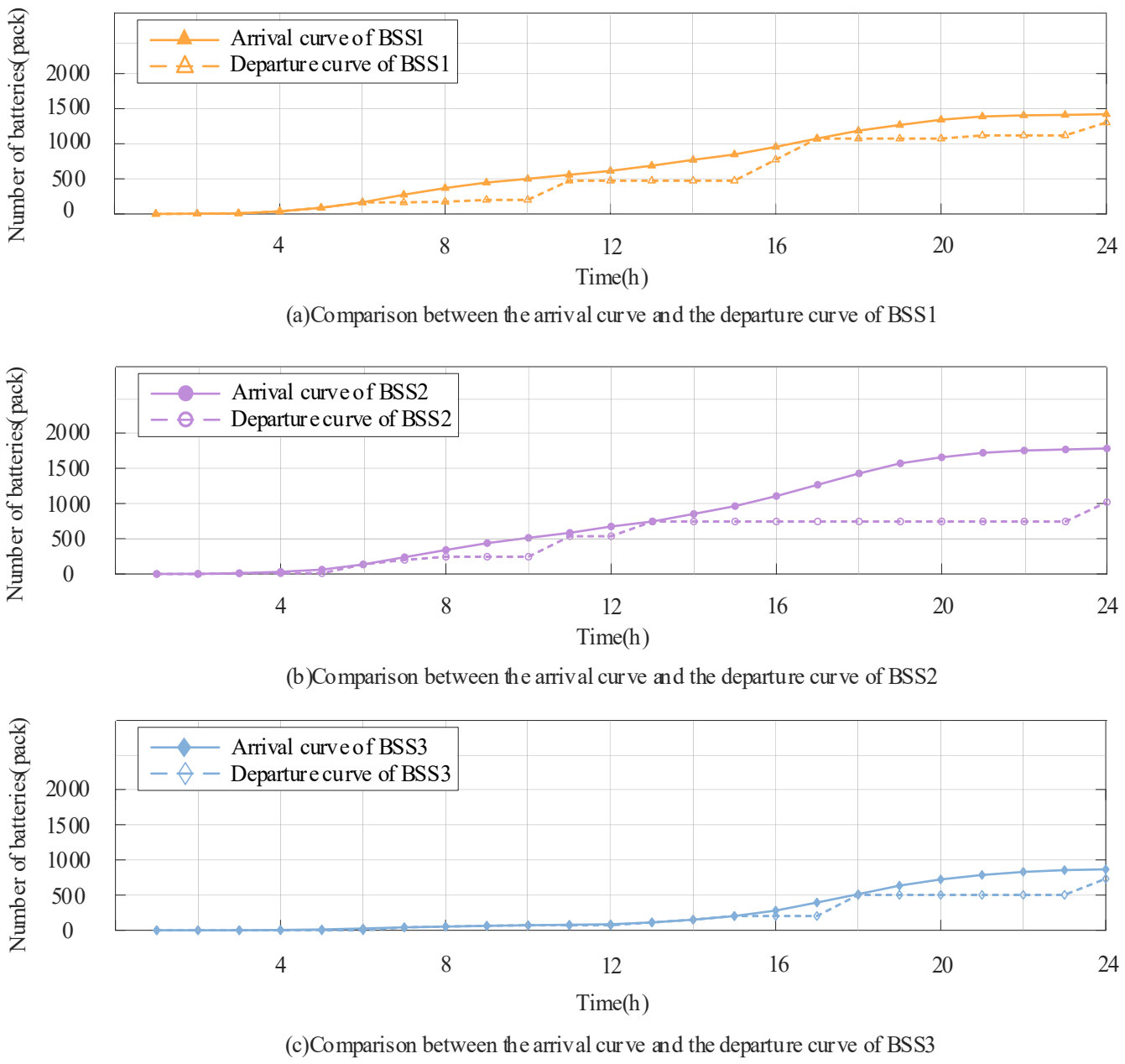

Fig. 16. Battery swapping results of all BSSs in the simplified benchmark model.

### 2) *Comparative Analysis Between the Proposed BCSS and a Model That Does Not Consider the End-of-Period Return Constraint on BSS Battery Inventory*

When constructing the BSS operational model, this paper introduces end-of-period return constraints on battery inventory, namely Equations (24) and (25). To demonstrate the specific role of these constraints in the model, this section simulates a BCSS without these constraints (hereinafter referred to as Scenario B) and systematically compares it with the proposed BCSS (hereinafter referred to as Scenario A).

The operational results of Scenario B are summarized in Table IV. Comparing them with Table II for Scenario A leads to the following conclusion: the presence or absence of the end-of-period return constraint significantly affects the simulation results. Specifically, Scenario A incurs higher operational costs for both the BCS and the logistics system, and its total operational revenue is lower than that of Scenario B. However, Scenario B exhibits abnormal DB inventory peaks at each BSS at the final time step. This phenomenon precisely reveals that in its pursuit of maximizing short-term revenue, Scenario B sacrifices the continuity and stability of system operations. This section will discuss this point in detail.

TABLE IV. OPERATIONAL RESULTS OF SCENARIO B

| $P_{\text{obj,tran}}$ | $P_{\text{obj,BSS}}$ | $P_{\text{obj,BCS}}$ | $P_{\text{obj,all}}$ |
|---|---|---|---|
| -1666.5 $ | 24414 $ | -13750.01 $ | 8997.49 $ |

Under these two scenarios, the most significant difference lies in the DB inventory at each BSS, the WB inventory at each BSS, and the DB inventory at the BCS at the final time step, as shown in Table V. At the end of the current operational cycle, each BSS in Scenario B has no WB inventory, and the BCS also stores no DBs or WBs. This implies that the model cannot provide sufficient WBs in a timely manner to meet the battery swap demand at the beginning of the next cycle, causing eVTOLs to wait for extended periods at each BSS. Even without traffic congestion, this inventory state already harbors the risk of operational failures; once congestion occurs, it is highly likely to lead to severe operational accidents and substantial economic losses. When the end-of-period return constraint is present, each BSS retains an appropriate amount of WB inventory, sufficient to support the battery swap demand at the beginning of the next operational cycle. At the same time, the remaining 258 DBs at the BCS can directly enter the charging process, allowing the BCS to avoid calling upon spare DBs and to charge without waiting for trucks to return DBs from the BSSs.

TABLE V. DIFFERENCES IN BATTERY INVENTORY BETWEEN SCENARIO A AND SCENARIO B

| *Scenario A* | | *Scenario B* | |
|---|---|---|---|
| ***Parameter*** | ***Value*** | ***Parameter*** | ***Value*** |
| $S^{1,T}_{\text{BSS,DB,stor}}$ (pack) | 230 | $S^{1,T}_{\text{BSS,DB,stor}}$ (pack) | 466 |
| $S^{2,T}_{\text{BSS,DB,stor}}$ (pack) | 230 | $S^{2,T}_{\text{BSS,DB,stor}}$ (pack) | 800 |
| $S^{3,T}_{\text{BSS,DB,stor}}$ (pack) | 230 | $S^{3,T}_{\text{BSS,DB,stor}}$ (pack) | 625 |
| $S^{1,T}_{\text{BSS,WB,stor}}$ (pack) | 170 | $S^{1,T}_{\text{BSS,WB,stor}}$ (pack) | 0 |
| $S^{2,T}_{\text{BSS,WB,stor}}$ (pack) | 170 | $S^{2,T}_{\text{BSS,WB,stor}}$ (pack) | 0 |
| $S^{3,T}_{\text{BSS,WB,stor}}$ (pack) | 170 | $S^{3,T}_{\text{BSS,WB,stor}}$ (pack) | 0 |
| $S^{1,T}_{\text{BCS,DB,stor}}$ (pack) | 258 | $S^{1,T}_{\text{BCS,DB,stor}}$ (pack) | 0 |
| $S^{1,T}_{\text{BCS,WB,stor}}$ (pack) | 0 | $S^{1,T}_{\text{BCS,WB,stor}}$ (pack) | 0 |

From Table IV, it can also be seen that the total operational revenue of Scenario B is USD 2,664.91 higher than that of Scenario A, of which USD 2,309.81 comes from the reduction in BCS operational costs. Further analysis of the BCS data for the two scenarios reveals that Scenario B calls upon only 820 spare DBs and supplies a total of 3,469 WBs to the logistics system, significantly lower than the corresponding figures of 958 and 3,979 in Scenario A. The main reason for this difference is that each BSS in Scenario B exhausts its initial WB inventory entirely over the operational cycle, alleviating the pressure on the BCS in terms of battery charging and spare DB calls, thereby reducing its operational costs. Furthermore, at the end of the operational cycle, the total battery inventory at each BSS in Scenario B is higher than that in Scenario A. While a lack of adequate battery reserves at a BSS is clearly unreasonable, excessive inventory also implies an increased potential safety risk.

*3) Comparative Analysis Between the Proposed BCSS and a Model That Does Not Consider the End-of-Period Return Constraint on Truck Battery Inventory*

This paper also imposes an end-of-period return constraint on truck battery inventory in the logistics system operational model, as shown in Equation (9). To demonstrate the effectiveness of this constraint, this section conducts a comparative analysis between a BCSS without the end-of-period return constraint (hereinafter referred to as Scenario C) and Scenario A, and elaborates on the impacts and advantages of retaining this constraint from three aspects: battery charging scheduling, inventory allocation, and logistics transportation.

The key operational results of Scenario C are shown in Table VI. Although its total operational revenue is USD 44.77 higher than that of Scenario A, this marginal advantage cannot mask its obvious deficiencies in practical operation. Regarding battery charging scheduling, a comparison of BCS operational results between the two scenarios shows that Scenario C uses three fewer spare DBs than Scenario A, suggesting that its BCS operational cost should theoretically be lower. However, the charging strategy of Scenario C is less reasonable than that of Scenario A, exhibiting an abnormal pattern of "charging more when electricity prices are high and charging less when prices are low," which results in a BCS operational cost that is USD 40.73 higher than that of Scenario A. For example, the electricity price at time step 6 is significantly higher than that at time step 7, yet Scenario C charges 210 and 283 DBs at these two time steps, respectively, in stark contrast to Scenario A's 196 and 300 DBs. Similar patterns can also be observed at time steps 12, 14-18, and 20-21.

TABLE VI. OPERATIONAL RESULTS OF SCENARIO C

| $P_{\text{obj,tran}}$ | $P_{\text{obj,BSS}}$ | $P_{\text{obj,BCS}}$ | $P_{\text{obj,all}}$ |
|---|---|---|---|
| -1936.1 $ | 24414 $ | -16100.55 $ | 6377.35 $ |

Regarding battery inventory allocation, the operational results of Scenario C are not conducive to the next cycle of operations. First, it should be noted that over the entire operational cycle, the total number of DBs supplied to the logistics system by all BSSs is the same between the two scenarios, the total number of WBs received is the same, and the total number of WBs supplied to the logistics system by the BCS is also the same. Second, after the conclusion of operations, the DB inventory at each BSS is 230 units, the WB inventory is 170 units, there is no WB inventory at the BCS, and no WBs are loaded on any of the four trucks. The difference lies in the fact that, under Scenario A, each truck carries only 100 DBs, whereas under Scenario C, Trucks 1 to 4 carry 213, 206, 300, and 236 DBs, respectively. On the surface, Scenario C's configuration appears more favorable for initiating a new operational cycle, as a large number of DBs are already loaded on trucks and can be sent directly back to the BCS at the very beginning of the new cycle. However, under Scenario A, the BCS already holds 258 DBs in inventory at the final moment, along with 300 DBs stored in chargers and ready for charging at any time. In contrast, under Scenario C, the BCS has no DBs at all. This implies that in the new cycle, Scenario A can commence charging operations earlier, without having to wait for trucks to return from BSSs as Scenario C does.

The scheduling results of the logistics system can explain the causes of the above phenomenon. Due to the lack of incentive from the end-of-period return constraint, Scenario C exhibits significantly lower truck transportation activity than Scenario A, with the four trucks being in transport for only 44 time steps over the operational cycle, compared to 47 time steps in Scenario A. This phenomenon is the fundamental reason why the BCS in Scenario C has no DB inventory and why the operational cost of the logistics system is USD 85.5 lower than that in Scenario A. In other words, by retaining the end-of-period return constraint on truck battery inventory, this paper obtains a more stable and sustainable BCSS at the cost of only USD 44.77 in operational revenue.

*4) Comparative Analysis Between the Proposed BCSS and a Model Considering Traffic Congestion*

Traffic congestion is inevitable in urban transportation. If trucks transporting WBs fail to arrive on time, the BCSS will face the risk of battery supply chain disruption. To enhance the operational resilience of the proposed BCSS under traffic congestion scenarios, this section imposes Constraint (40) on the BSSs and analyzes the impact of this constraint on the overall system operation. Simulation results show that the proposed BCSS can maximize operational revenue as much as possible while ensuring the continuity of battery swap services.

Constraint (40) enables the BCSS to maintain the operation of battery swap services even when traffic congestion persists for $\theta$ hours. That is, the sum of $S_{\text{BSS,WB,storage}}^{n,\text{ini}}$ of BSS $n$ and the cumulative number of WBs received by that BSS up to time step $t$ shall not be less than the cumulative number of WBs used by that BSS for battery swapping up to time step $t+\theta$ .

$$S^{n,\text{ini}}_{\text{BSS,WB,stor}} + \sum_{h=1}^{t} N^{n,t}_{\text{BSS,WB,rec}} \ge D^{n,t+\theta}, \forall n \in G_{\text{BSS}}, t \in \mathcal{T}-\theta \quad (40)$$

This paper defines three simulation scenarios. The BCSS without considering traffic congestion described above serves as the standard model, referred to as Scenario A. The BCSS considering one hour of traffic congestion ($\theta = 1$) is referred to as Scenario D. The BCSS considering two hours of traffic congestion ($\theta = 2$) is referred to as Scenario E. Fig. 17 shows the total WB inventory across all BSSs at each time step for the three scenarios. When traffic congestion is taken into account in the BCSS, the total WB inventory across all BSSs in Scenarios D and E is never lower than that in Scenario A. In terms of overall trend, the longer the congestion duration, the higher the total WB inventory at each time step, with exceptions occurring at time steps 15, 16, and 19. Since Scenario E provides far more services than Scenario D at these time steps, its total WB inventory temporarily falls below that of Scenario D. The reason for the above overall trend is that the BCSS must maintain a high level of WB inventory at the BSSs over extended periods in advance to cope with occasional traffic congestion, thereby avoiding interruption of battery swap services due to a lack of WBs.

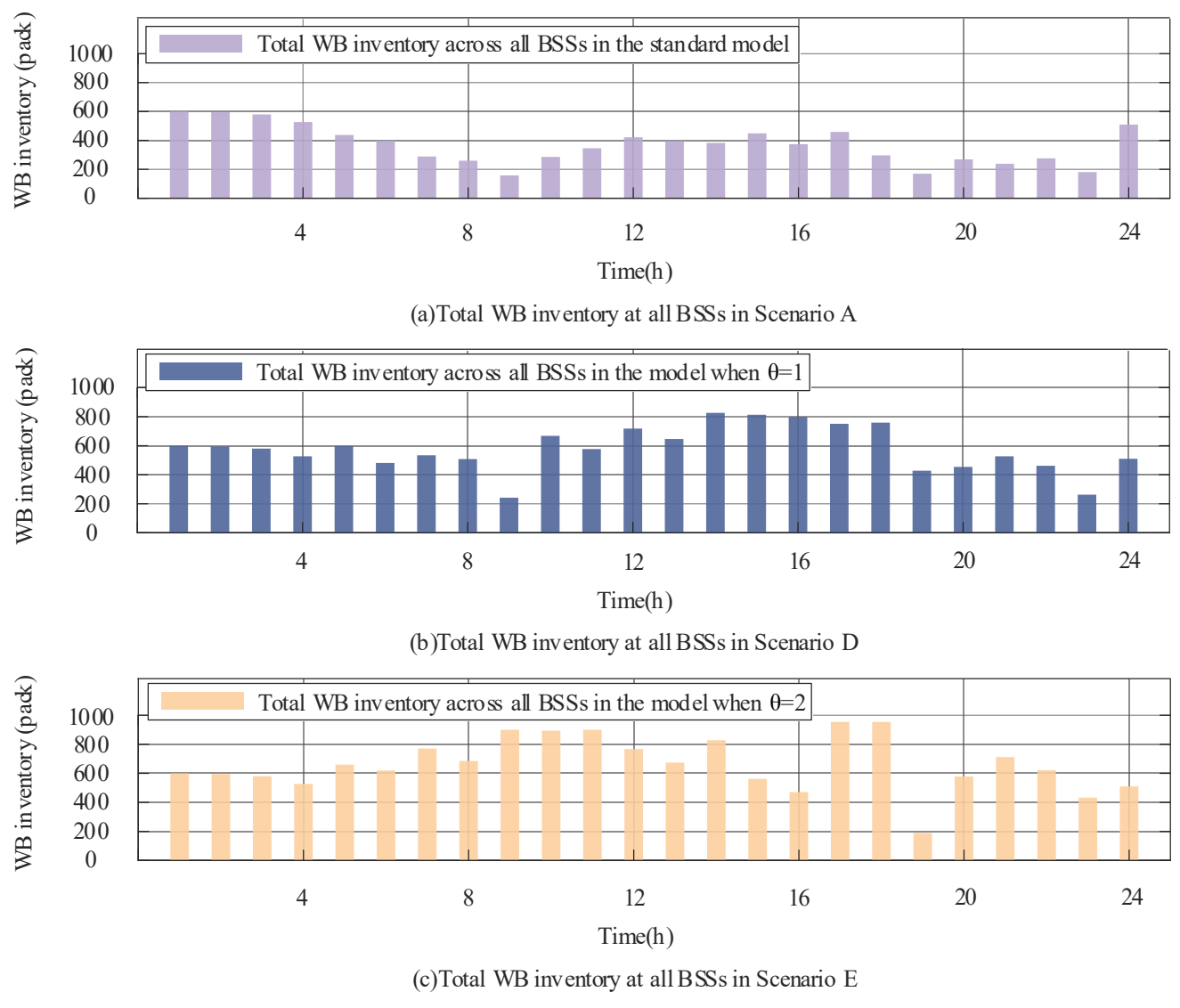

Fig. 17. Sum of WB inventory at all BSSs per time step for three scenarios.

Table VII also presents the operational results of Scenarios D and E. It can be seen that the results of the logistics system and BSSs are essentially consistent across the three scenarios. However, as the congestion duration increases, the BCS operational cost continues to rise, ultimately leading to a continuous decline in the overall operational revenue of the BCSS. Analysis of their BCS operational data reveals that Scenarios D and E require 1,184 and 1,322 spare DBs, respectively, far exceeding the 958 spare DBs required in Scenario A. The main reason is that, to meet the excessively high WB demand of the BSSs, the BCS must dispatch a large number of spare DBs within the first six time steps of the operational cycle. While it is clearly difficult to meet subsequent operational demands by relying excessively on the gradual return of DBs by the fleet for subsequent charging, this inevitably incurs high dispatch costs.

Furthermore, the charging strategies of Scenarios D and E also differ from that of Scenario A. As shown in Fig. 18, Scenarios D and E need to commence substantial charging during the early operational period when electricity prices are relatively high. For example, at time steps 2, 6, and 8, the BCS in Scenario A removes 0, 196, and 262 WBs from the chargers, respectively; at the same time steps, Scenario D removes 161, 232, and 295 WBs, respectively, while Scenario E removes 222, 300, and 300 WBs, respectively. However, because Scenarios D and E charge on a large scale in the early stage of operations, their WB reserves in the later stage satisfy subsequent battery swap demands ahead of schedule, eliminating the need for further large-scale charging. In contrast, Scenario A relies more heavily on DBs returned by trucks from the BSSs, which prevents it from meeting BSS demands in advance and forces it to charge more batteries at time steps with even higher electricity prices. For example, at time steps 20 and 21, Scenario A removes 298 and 215 WBs, respectively, whereas Scenario D removes only 194 and 170 WBs, respectively, and Scenario E only needs to remove 170 WBs at time step 20.

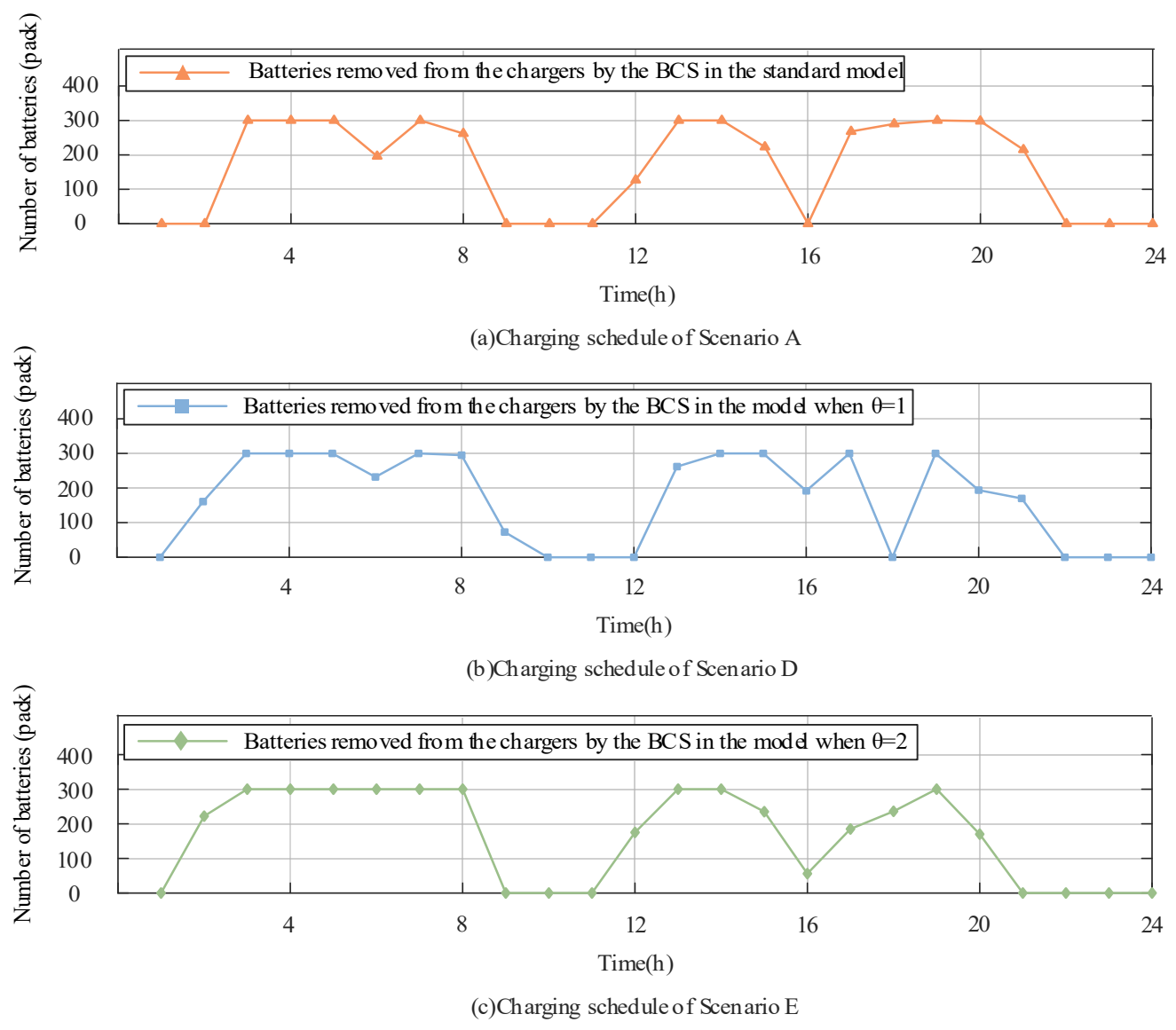

Fig. 18. Number of WB taken from charging piles per BCS for three scenarios.

TABLE VII. OPERATIONAL RESULTS OF SCENARIOS D AND E

| *Scenario* | $P_{\text{obj,tran}}$ | $P_{\text{obj,BSS}}$ | $P_{\text{obj,BCS}}$ | $P_{\text{obj,all}}$ |
|---|---|---|---|---|
| D | -2031.6 \$ | 24414 \$ | -17119.1 \$ | 5263.3 \$ |
| E | -2021.6 \$ | 24414 \$ | -17731.67 \$ | 4660.73 \$ |

## VI. CONCLUSION

This paper addresses the fast energy replenishment problem for eVTOL batteries by proposing a battery charging and swapping system model based on CLSC and TSN. With the objective of maximizing operational revenue, all members within the system are optimally managed. Finally, the feasibility of the proposed model is verified through simulation, and the following conclusions are drawn:

(1) The scheduling optimization strategy developed by the BCSS in this paper achieves efficient operation and coordination among all members. In terms of battery transportation, the logistics system model dynamically schedules trucks based on actual battery swap demand and the battery inventory status of each member, achieving optimal allocation of transportation and battery resources while helping the BCS reduce operational costs. In terms of battery swap operations, the BSS makes the most reasonable battery swap service decision at each time step by comprehensively considering factors such as battery inventory, battery swap service quality, and WB supply quantity. In terms of energy management, the BCS, subject to satisfying the demand of the BSSs, provides a stable battery supply for the entire system while minimizing operational costs through means such as appropriately leasing spare DBs and finely regulating the charging schedule.

(2) In comparison with the simplified benchmark model, the proposed BCSS demonstrates strong potential for sustainable development, prioritizing the guarantee of battery swap service quality even at the cost of sacrificing a portion of operational revenue. In comparison with a BCSS without end-of-period return constraints, the proposed BCSS exhibits significant advantages in ensuring the continuity of system operations, avoiding imbalanced battery distribution that would impede the smooth startup of the next operational cycle even if it means forgoing a portion of operational revenue in the current period. By introducing appropriate constraints, the proposed BCSS shows good resilience in the face of traffic congestion, ensuring the continuity of battery swap services while still achieving considerable positive revenue.

(3) Future research can be carried out in depth from the following three aspects based on this paper. First, compress the time step size to characterize the model operation process more finely, thereby improving the timeliness and accuracy of decision-making. Second, introduce renewable energy sources such as wind and solar power as supplemental power supplies to enhance the flexibility of system power supply. Third, further explore the potential of batteries to facilitate their participation in ancillary services such as grid demand response and frequency regulation.

(4) In addition to Shanghai, major global economically developed cities such as Berlin, Paris, and Tokyo are also actively advancing the development of eVTOLs. Therefore, the BCSS constructed in this paper possesses certain engineering value and broad application scenarios. However, when implementing the model proposed in this paper in a specific city, targeted modifications to the model must still be made in accordance with local characteristics. For instance, a more complex traffic network model that better reflects actual road conditions could be constructed; the uncertainty of passenger travel demand could also be incorporated into the BCSS; and even the transfer and connection issues for passengers between eVTOLs and ground public transportation (such as subways) could be further considered.